\documentclass[aps,prl,twocolumn,superscriptaddress,]{revtex4-2}
\usepackage{amsmath}
\usepackage{amsfonts}
\usepackage{graphicx}
\usepackage{color}
\usepackage{dsfont}
\usepackage{verbatim}
\usepackage{mathtools}
\usepackage[normalem]{ulem}
\usepackage{comment}
\usepackage{stackrel}
\usepackage{bm}
\usepackage[pdftex]{hyperref}
\hypersetup{colorlinks = true, urlcolor = black, linkcolor = black, citecolor = black}

\definecolor{myblue}{rgb}{.93, .93, 1}

\newcommand{\bsub}{\begin{subequations}}
	\newcommand{\esub}{\end{subequations}}

\newcommand{\vex}[1]{\bm{\mathrm{#1}}}

\begin{document}

	\title{Kondo lattice model in magic-angle twisted bilayer graphene}
	
	\author{Yang-Zhi~Chou}\email{yzchou@umd.edu}
	\affiliation{Condensed Matter Theory Center and Joint Quantum Institute, Department of Physics, University of Maryland, College Park, Maryland 20742, USA}

	\author{Sankar Das~Sarma}
	\affiliation{Condensed Matter Theory Center and Joint Quantum Institute, Department of Physics, University of Maryland, College Park, Maryland 20742, USA}	
	\date{\today}
	
	\begin{abstract}
	We systematically study emergent Kondo lattice models from magic-angle twisted bilayer graphene using the topological heavy fermion representation. At the commensurate fillings, we demonstrate a series of symmetric strongly correlated metallic states driven by the hybridization between a triangular lattice of $SU(8)$ local moments and delocalized fermions.
	In particular, a (fragile) topological Dirac Kondo semimetal can be realized, providing a potential explanation for the symmetry-preserving correlated state at $\nu=0$. We further investigate the stability of the Dirac Kondo semimetal by constructing a quantum phase diagram showing the interplay between Kondo hybridization and magnetic correlation. The destruction of Kondo hybridization suggests that the magic-angle twisted bilayer graphene may be on the verge of a solid-state quantum simulator for novel magnetic orders on a triangular lattice. Experimental implications are also discussed. 
\end{abstract}

\maketitle

\textit{Introduction.---}
Magic-angle twisted bilayer graphene (MATBG) \cite{Cao2018_tbg1,Cao2018_tbg2} has been a promising platform to study strongly correlated phases because the nearly flat bandwidth \cite{BM_model2011} effectively enhances many-body interactions. Recent experiments \cite{Cao2018_tbg1,Cao2018_tbg2,Yankowitz2019,Kerelsky2019,Lu2019,Sharpe2019,Serlin2020,Jiang2019,Xie2019_spectroscopic,Polshyn2019,Cao2020PRL,Park2021flavour,Choi2019electronic,Zondiner2020cascade,Arora2020superconductivity,Wong2020cascade} have shown abundant phenomena such as correlated insulators, superconductivity, quantum anomalous Hall effect, and flavor polarization. One of the unsettled experimental issues is the seeming contradiction between transport and STM measurements at the charge neutrality point ($\nu=0$). Several transport experiments \cite{Cao2018_tbg1,Cao2018_tbg2,Yankowitz2019,Zondiner2020cascade} found semimetallic behavior at $\nu=0$, which can be described by the noninteracting Bistritzer-MacDonald (BM) model \cite{BM_model2011}. However, STM measurements \cite{Xie2019_spectroscopic,Wong2020cascade,Choi2019electronic,Jiang2019} observed strong local correlations at $\nu=0$, indicating significant interactions. Thus, it is natural to ask if a potential mechanism exists explaining both transport and STM experiments.

Studying interaction-driven phenomena in MATBG is technically challenging because the two flat bands belong to the fragile topological states protected by a $\mathcal{C}_{2z}\mathcal{T}$ symmetry \cite{Song2019,Po2019,Ahn2019}, implying the absence of a lattice model description for the flat bands. Recently, Song and Bernevig proposed a novel topological heavy fermion (THF) representation \cite{Song2022}, reconstructing the MATBG flat bands by coupling localized orbitals ($f$ fermions) and delocalized topological conduction bands ($c$ fermions). Notably, the localized $f$ fermions can be viewed as the zeroth pseudo-Landau levels located at the AA stacking registries \cite{Liu2019_Pseudo_LL,Shi2022HF_Rep}. Song and Bernevig further demonstrated that this THF representation is advantageous for studying interaction-driven phases because the one-shot Hartree-Fock results qualitatively capture the essence of self-consistent Hartree-Fock calculations \cite{Song2022}, suggesting a good starting point for exploring strongly correlated physics in MATBG.

\begin{figure}[t!]
	\includegraphics[width=0.475\textwidth]{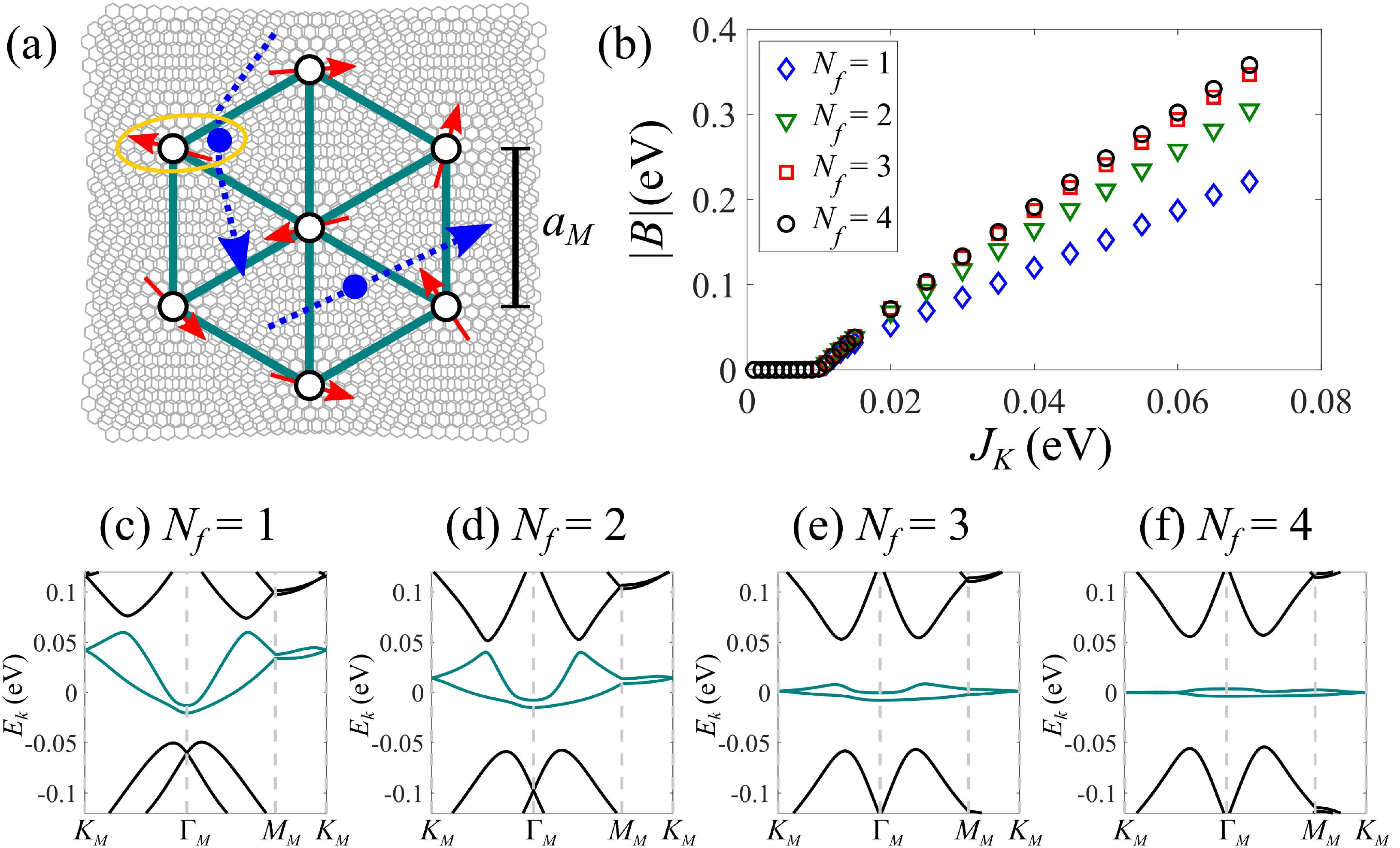}
	\caption{Kondo lattice model for MATBG. (a) The real-space structure. Local moments (red arrows) are located at the AA stacking registries and form a triangular lattice with a lattice constant $a_M$. The delocalized $c$ fermions (blue dots) interact with these local moments. (b) The Kondo hybridization amplitude as a function of $J_K$ with different values of $N_f$. The results are obtained by solving Eqs.~(\ref{Eq:B}), (\ref{Eq:Nf}), and (\ref{Eq:Nc}) with a $30\times30$ momentum mesh in each mBZ. (c)-(f) Quasiparticle dispersion of $\eta=+1$ for $J_K=0.03$ eV with different $N_f$. $(\nu_f,\nu_c)=(N_f-4,0)$ in these plots. The mini bands are highlighted with green color.}
	\label{Fig:Fig1}
\end{figure}

In this Letter, we construct and study Kondo lattice models for MATBG using the newly developed THF model \cite{Song2022,Shi2022HF_Rep}. The $f$ orbitals can be viewed as $SU(8)$ local moments when the onsite Hubbard interaction is sufficiently large. We show that a series of symmetry-preserving correlated metallic states arise naturally from the hybridization between the $SU(8)$ local moments (spin, valley, and orbital) and the delocalized $c$ fermion. In particular, a fragile topological Dirac Kondo semimetal is realized at $\nu=0$, providing a potential resolution for both the transport and STM measurements at $\nu=0$. This prediction is distinct from the existing Hartree-Fock studies \cite{Xie2020_Correlated_Insulator,Zhang2020_Correlated_ins,Bultinck2020,Liu2021_Th_correlated_ins,TBG_III,TBG_IV,TBG_V,Wagner2022,Song2022,Shi2022HF_Rep}. To examine the stability of the Dirac Kondo semimetal state, we consider Heisenberg interaction among the nearest-neighbor local moments and construct a quantum phase diagram. One interesting implication is that MATBG might be on the verge of a $SU(8)$ Heisenberg model on a triangular lattice, paving an unprecedented way for studying exotic magnetic phases in solid-state systems.

\textit{Model.---} We are interested in the MATBG low-energy bands with strong correlations. The single-particle dispersion is well described by the BM model \cite{BM_model2011}. Here, we adopt a new alternative strategy using the THF representation \cite{Song2022,Shi2022HF_Rep}. The main idea is that the low-energy MATBG bands can be reconstructed by coupling localized orbitals (at zero energy) and delocalized topological bands. The localized orbitals ($f$ fermions) live on a triangular lattice with the lattice constant $a_M$ being the moir\'e period as shown in Fig~\ref{Fig:Fig1}(a); the delocalized bands ($c$ fermions) travel in continuous space, described by \cite{Song2022}
\begin{align}\label{Eq:H_0_c}
	\hat{H}_{0,c}\!=\!\sum_{\eta,s,a,a'}\sum_{\vex{q}}\!h^{(\eta)}_{aa'}(\vex{q})c^{\dagger}_{\vex{q},a,\eta,s}c_{\vex{q},a',\eta,s},
\end{align} 
where $\hat{h}^{(\eta)}(\vex{q})$ is a $4\times 4$ matrix dictating the delocalized bands \cite{SM}, $c_{\vex{q},a,\eta,s}$ is annihilation operator for the delocalized bands with valley $\eta$, spin $s$, orbital $a$ ($a=1,2,3,4$), and is a wavevector $\vex{q}$. Since $c$ fermions can travel in the continuous space freely (i.e., not trapped by the triangular lattice points), they can carry wavevectors outside of the first moir\'e Brillouin zone (mBZ) defined by the triangular superlattice. The hybridization between $c$ and $f$ fermions is described by \cite{Song2022}
\begin{align}
	\label{Eq:H_0_cf}\hat{H}_{0,cf}\!\!=\!\!\!\!\!\sum_{\eta,s,\alpha,a}\sum_{\vex{G}}\!\sum_{\vex{k}\in \text{mBZ}}\!\!\!\left[V^{(\eta)}_{\alpha a}(\vex{k}\!+\!\vex{G})f^{\dagger}_{\vex{k},\alpha,\eta,s}c_{\vex{k}+\vex{G},a,\eta,s}\!+\!\text{H.c.}
	\right]\!,
\end{align}
where $V^{(\eta)}_{\alpha a}(\vex{q})$ is a $2\times 4$ matrix \cite{SM}, $f_{\vex{k},\alpha,\eta,s}$ is the annihilation operator for the localized orbital with orbital $\alpha$ ($\alpha=1,2$), valley $\eta$, spin $s$, and wavevector $\vex{k}$ defined in the first mBZ, and $\vex{G}$ is the reciprocal lattice vector of the triangular superlattice. The characteristic single-particle hybridization is $\gamma=-24.75$ meV, and $|\gamma|$ controls the distance between the low-energy and remote bands \cite{Song2022}.
Equation~(\ref{Eq:H_0_cf}) conserves only the crystal momentum. Thus, the $f$ fermion with wavevector $\vex{k}$ couples to all the $c$ fermions with wavevectors $\vex{k}+\vex{G}$ for all the allowed $\vex{G}$. In our calculations, we consider a finite number ($N_G$) of reciprocal lattice vectors. $N_G=37$ is used for all the results.

In addition to the single-particle part, the Coulomb interaction can be projected into the THF basis \cite{CL}, and the interacting part of the Hamiltonian is given by $\hat{H}_I=\hat{H}_U+\hat{H}_V+\hat{H}_W+\hat{H}_J$ \cite{Song2022}, where
\begin{align}
	\label{Eq:H_U}\hat{H}_{U}=\frac{U}{2}\sum_{\vex{R}}\left(\hat{\rho}_{\vex{R}}^f-4\right)^2
\end{align}
describes the onsite repulsive interactions ($U>0$) among the $f$ fermions \cite{U2}, $\hat{H}_V$ denotes the Coulomb interaction between $c$ fermions. $\hat{H}_W$ is the density-density interaction between $f$ and $c$ fermions (similar to the interaction in the Falicov-Kimball model), and $\hat{H}_J$ is a $U(4)$ Hund's rule coupling. 
In Eq.~(\ref{Eq:H_U}), the subtraction of $4$ incorporates the effect of the ionic charge background.

To study the interacting MATBG, we make a few simplifications \cite{CL}. We ignore the $\hat{H}_V$ term because it mostly renormalize the $c$ fermion dispersion (e.g., band velocity and chemical potential) but unlikely induces any qualitatively change in the results. We also neglect the $\hat{H}_W$ term as it primarily gives shifts of chemical potentials. 
While $\hat{H}_J$ is crucial for stabilizing correlated insulating states in Hartree-Fock calculations \cite{Song2022}, it is irrelevant to the symmetric Kondo correlated metals within the mean-field treatment as we discussed in Supplementary Material \cite{SM}. Since we focus only on the Kondo-driven phases, the minimal interacting model for MATBG is described by $\hat{H}_{0,c}+\hat{H}_{0,cf}+\hat{H}_U$ [Eqs.~(\ref{Eq:H_0_c}), (\ref{Eq:H_0_cf}), and (\ref{Eq:H_U})].

\textit{Kondo lattice and heavy fermion phase.---} The Hamiltonian $\hat{H}_{0,c}+\hat{H}_{0,cf}+\hat{H}_U$ can be viewed as a periodic Anderson model \cite{Coleman2015introduction,Hewson1997kondo} for MATBG. When $U$ is sufficiently strong (i.e., $U\gg|\gamma|$), the local occupation number of $f$ fermions ($N_f$) is frozen at each site. In such a situation, the $cf$ hybridization ($\hat{H}_{0,cf}$) is inert, and a Kondo coupling emerges at the second order of $\hat{H}_{0,cf}$ \cite{Coleman2015introduction,Hewson1997kondo} (see supplementary material \cite{SM}),
\begin{align}
	\nonumber\hat{H}_K\!=&\frac{J_K}{\mathcal{N}_k\gamma^2}\!\sum_{\substack{\alpha,\eta,a,s\\ \alpha',\eta',a',s'}}\sum_{\vex{G},\vex{G}'}\sum_{\vex{k},\vex{k}'}\sum_{\vex{R}}\left[V^{(\eta')}_{\alpha'a'}(\vex{k}'\!+\!\vex{G}')\right]^*V^{(\eta)}_{\alpha a}(\vex{k}\!+\!\vex{G})\\
	\label{Eq:H_K}&\!\times\! e^{i(\vex{k}-\vex{k}')\cdot\vex{R}}\!\!:\!\!f^{\dagger}_{\vex{R},\alpha,\eta,s}f_{\vex{R},\alpha',\eta',s'}\!::\!c^{\dagger}_{\vex{k}'+\vex{G}',a',\eta',s'}c_{\vex{k}+\vex{G},a,\eta,s}\!\!:\!\hat{\mathcal{P}}_{N_f},
\end{align}
where $:\!A\!:$ denotes the normal order of $A$ \cite{NO}, and $\hat{\mathcal{P}}_{N_f}$ is the projection operator onto the subspace with exactly $N_f$ localized $f$ fermions per site. 
In the above expression, the Kondo coupling has a nontrivial momentum dependence. In a limit that $\vex{k}+\vex{G}\rightarrow0$ and $\vex{k}'+\vex{G}'\rightarrow 0$, $\left[V^{(\eta')}_{\alpha'a'}(0)\right]^*V^{(\eta)}_{\alpha a}(0)$ is reduced to $\gamma^2\delta_{a',\alpha'}\delta_{a,\alpha}$, and Eq.~(\ref{Eq:H_K}) becomes to a $SU(8)$ Coqblin-Schrieffer coupling \cite{Coqblin1969,Coleman2015introduction,Hewson1997kondo} with a coupling constant $J_K$. Using $U=57.95$ meV (the same valued used in Ref.~\cite{Song2022}), we obtain $J_K\approx 42.28$ meV, indicating that $J_K$ cannot be ignored. In addition, the value of $J_K$ is insensitive to $N_f$ \cite{SM}. We emphasize that $\hat{H}_K$ and $\hat{H}_J$ act on different orbital subspaces of $c$ fermions, so they should be treated separately.

To study the Kondo lattice model $\hat{H}_{0,c}+\hat{H}_K$, we employ the Read-Newns decoupling \cite{Read1983solution,Coleman2015introduction,Hewson1997kondo} (i.e., hybridization decoupling). The main results are summarized in the main text, and the derivations can be found in Supplemental Material \cite{SM}. In the mean-field theory, the $\hat{H}_K$ is replaced by
\begin{align}
	\nonumber\hat{H}_K\!\rightarrow&\!\sum_{\substack{\eta,s\\
			\alpha,a}}\sum_{\vex{G}}\!\sum_{\vex{k}\in\text{mBZ}}\!\!\left[\!\frac{B}{\gamma}V^{(\eta)}_{\alpha a}\!(\vex{k}\!+\!\vex{G})f^{\dagger}_{\vex{k},\alpha,\eta,s}c_{\vex{k}+\vex{G},a,\eta,s}\!\!+\!\text{H.c.}
	\right]\\
	&+\sum_{\vex{k}\in\text{mBZ}}\frac{|B|^2}{J_K},
\end{align}
where the Kondo hybridization
\begin{align}
	\label{Eq:B}B\!=\!-\frac{J_K}{\mathcal{N}_k}\!\sum_{\vex{G}}\sum_{\vex{k}\in\text{mBZ}}\sum_{\alpha,\eta,s}\!\frac{V^{(\eta)}_{\alpha a}(\vex{k}\!+\!\vex{G})}{\gamma}\!\left\langle f^{\dagger}_{\vex{k},\alpha,\eta,s}c_{\vex{k}+\vex{G},a,\eta,s}\right\rangle,
\end{align}
and $\mathcal{N}_k$ is the number of $\vex{k}$'s in the first mBZ (equivalent to the number of lattice points). This mean-field decoupling is asymptotically valid for a $SU(N)$ system with $N\gg1$ \cite{Coleman2015introduction,Hewson1997kondo}. We use the same mean-field decoupling for $\hat{H}_J$ and find that $\hat{H}_J$ vanishes exactly as long as the states are symmetric in valleys and orbitals \cite{SM}. This technical observation indicates that $\hat{H}_J$ is not essential for the Kondo lattice problems without valley/orbital symmetry breaking.

Finally, we need to impose a local constraint such that each site contains exactly $N_f$ $f$ fermions. In the mean-field treatment,
\begin{align}
	\label{Eq:Nf}\frac{1}{\mathcal{N}_k}\sum_{\vex{k}\in \text{mBZ}}\sum_{\alpha,\eta,s}\left\langle f^{\dagger}_{\vex{k},\alpha,\eta,s}f_{\vex{k},\alpha,\eta,s}\right\rangle=N_f.
\end{align}
Similarly, we can compute the number of $c$ fermions per site with a finite $N_G$ (the number of mBZs included),
\begin{align}
	\label{Eq:Nc}\frac{1}{\mathcal{N}_k}\sum_{\vex{G}}\sum_{\vex{k}\in \text{mBZ}}\sum_{a,\eta,s}\left\langle c^{\dagger}_{\vex{k}+\vex{G},a,\eta,s}c_{\vex{k}+\vex{G},a,\eta,s}\right\rangle=N_c.
\end{align}
The total filling is determined by $\nu=\nu_f+\nu_c$, where $\nu_f=N_f-4$ and $\nu_c=N_c-8N_G$. Assuming that the interaction effect is dominated by the $f$ fermions \cite{Song2022,Shi2022HF_Rep}, we treat the integer fillings as $\nu=\nu_f$ and $\nu_c=0$. For a non-integer filling, one needs to compute the zero-point energy (i.e., contributions from $\hat{H}_V$ and $\hat{H}_W$ \cite{Song2022}, which we ignore) to determine the precise $\nu_f$ and $\nu_c$. 

The mean-field action in the imaginary-time path integral is given by \cite{SM}
\begin{align}
	\nonumber\mathcal{S}_{\text{MF}}\!=&\frac{1}{\beta}\!\sum_{\omega_n}\!\sum_{\eta,s}\!\sum_{\vex{k}\in \text{mBZ}}\!\!\!\hat{\bar{\Psi}}^{\eta,s}_{\omega_n,\vex{k}}\!\left[-i\omega_n\!\!+\!\!\hat{\mathcal{H}}^{\eta,s}\!(\vex{k};\!B,\!\mu_c,\!\mu_f)\right]\!\!\hat{\Psi}^{\eta,s}_{\omega_n,\vex{k}}\\
	\label{Eq:S_MF_KL}&+\beta\mathcal{N}_k\left(\frac{\mathcal{B}^2}{J_K}\!+\!\mu_f N_f\!+\!\mu_cN_c\right),
\end{align}
where $\hat{\mathcal{H}}^{\eta,s}$ is a $\left(2+4N_G\right)\times\left(2+4N_G\right)$ matrix, $\hat{\Psi}^{\eta,s}_{\omega_n,\vex{k}}$ is a $(2+4N_G)$-component field made of $f_{\vex{k},\alpha,\eta,s}$ as well as $c_{\vex{k}+\vex{G},a,\eta,s}$, and $\mu_f$ ($\mu_c$) is the chemical potential for $f$ ($c$) fermions. We can straightforwardly show that the self-consistent equations given by Eqs.~(\ref{Eq:B}), (\ref{Eq:Nf}), and (\ref{Eq:Nc}) are equivalent to the saddle-point equations. Within the saddle-point approximation \cite{Coleman2015introduction},
the mean-field energy per site can be derived by integrating out the fermionic fields and taking the zero-temperature limit:
\begin{align}
	\nonumber\frac{E_{\text{MF}}}{\mathcal{N}_k}=&\frac{1}{\mathcal{N}_k}\sum_{\vex{k}\in \text{mBZ}}\sum_{\eta,s}\!\sum_{b=1}^{2+4N_G}\!\mathcal{E}^{\eta,s}_b(\vex{k})\Theta(-\mathcal{E}^{\eta,s}_b(\vex{k}))\\
	\label{Eq:E_MF}&+\left(\frac{|B|^2}{J_K}+\mu_f N_f+\mu_cN_c\right).
\end{align}
where $\mathcal{E}^{\eta,s}_b(\vex{k})$ is the eigenvalue of $\hat{\mathcal{H}}^{\eta,s}(\vex{k};B,\mu_c,\mu_f)$ and $\Theta(x)$ is the Heaviside function. 

To obtain the mean-field ground state, we numerically solve the self-consistent equations given by Eqs.~(\ref{Eq:B})-(\ref{Eq:Nc}) \cite{SM}. Then, we compute the $E_{\text{MF}}$ [Eq.~(\ref{Eq:E_MF})] of all the solutions. The ground state is the solution with the minimal $E_{\text{MF}}$. For $\nu_f=\nu_c=0$, $\mu_c$ and $\mu_f$ are exactly zero, so the self-consistent procedure can be greatly simplified. For general cases, one has to choose the correct values of $\mu_f$ and $\mu_c$ so that Eqs.~(\ref{Eq:Nf}) and (\ref{Eq:Nc}) are satisfied. Remarkably, for $N_f\neq 4$, $\mu_f$ obtained from solving self-consistent equations can be significantly different from the local chemical potential for deriving $\hat{H}_K$ [Eq.~\ref{Eq:H_K}], suggesting that the local chemical potential is strongly renormalized. We choose a gauge corresponding to $B<0$, and the results are not affected by such a choice \cite{Coleman2015introduction}. The detailed numerical procedures are discussed in Supplementary Material \cite{SM}.

\begin{figure}[t!]
	\includegraphics[width=0.45\textwidth]{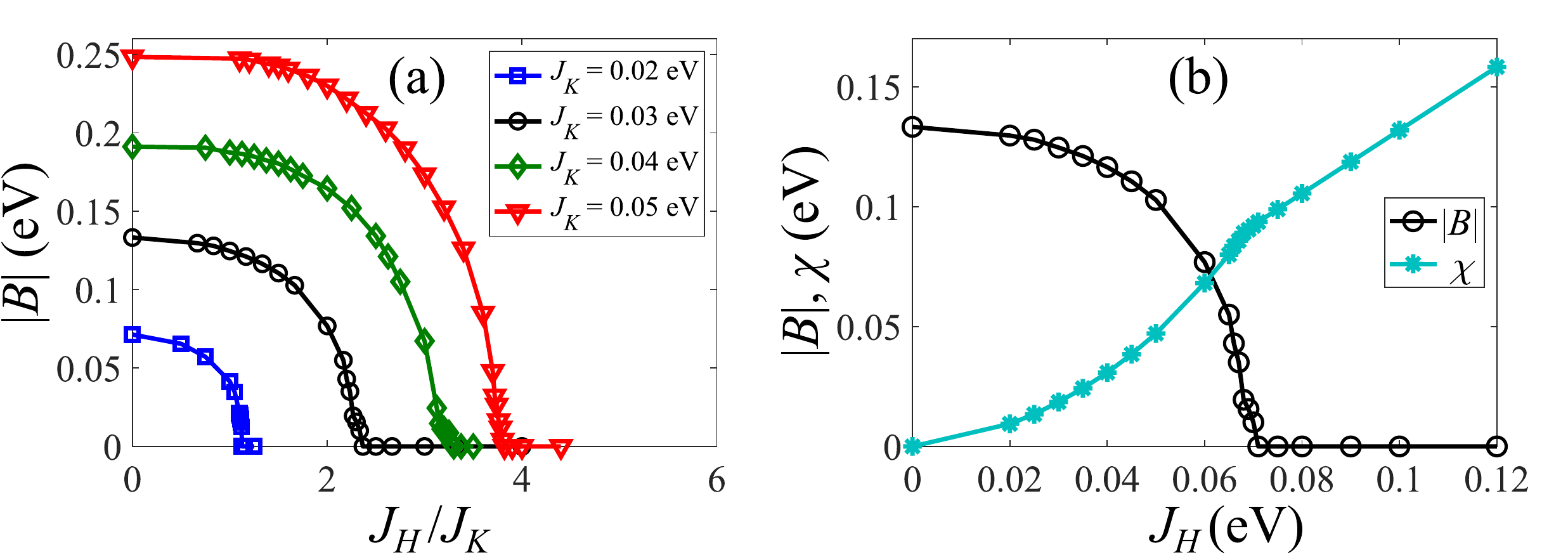}
	\caption{Hybridization amplitude and magnetic correlation in Kondo-Heisenberg model with $N_f=4$. (a) $|B|$ as a function of $J_H/J_K$ for different values of $J_K$. (b) $|B|$ and $\chi$ as functions of $J_H$ for $J_K=0.03$ eV.}
	\label{Fig:Fig2}
\end{figure}

Now, we construct a phase diagram for the Kondo hybridization formation.
In Fig.~\ref{Fig:Fig1}(b), we plot $|B|$ as a function of $J_K$ for different $N_f$ with $\nu_c=0$. The results suggest a threshold around $J_K= 0.01$ eV, above which $|B|$ becomes finite. For $J_K< 0.01$ eV, we conclude that $|B|\rightarrow 0$ based on finite-size analysis \cite{SM}. The filling factor $\nu$ can also modify $|B|$. For $J_K>0.01$ eV, we find that $\nu=0$ ($N_f=4$) yields the largest $|B|$, and $\nu=-3$ ($N_f=1$) gives the smallest $|B|$ as plotted in Fig.~\ref{Fig:Fig1}(b). Meanwhile, the quasiparticle dispersions along several line cuts ($K_M-\Gamma_M-M_M-K_M$) in Fig.~\ref{Fig:Fig1}(c)-(f) show that the mini bandwidth becomes wider for a larger $|\nu|$, and no spectral gap in the low-energy bands for all the cases, suggesting symmetry-preserving correlated metallic states at all the integer fillings. The broadening of quasiparticle low-energy bands for the nonzero integer fillings is a manifestation of strong correlation \cite{SM}. We also find interaction-driven phase transitions in the quasiparticle bands of $\nu=-3$ and $\nu=-2$ \cite{SM}.
Intriguingly, the $\nu=0$ case with $|B|>3.7$ meV gives low-energy bands very similar to the single-particle bands of MATBG \cite{KSM}, i.e., nearly flat isolated Dirac bands with the same bandwidth; the crucial differences are in the remote bands, particularly, the gaps between low-energy and the remote bands are proportional to $|B|$. We also point out that such low-energy bands also belong to the fragile topological bands \cite{Song2019,Po2019,Ahn2019} because of the topological equivalence to the single-particle bands. However, the physical origin here is due to strong correlations. Thus, we conclude that the Kondo lattice model realizes a fragile topological Dirac Kondo semimetal at $\nu=0$.

So far, we have discussed the results based on the Kondo lattice model, where $U$ is treated nonperturbatively. It is interesting to inspect the results in the perturbative regime. Following the ideas in Ref.~\cite{Edwards1988transition,Hewson1997kondo,Yamada1986fermi}, we treat the effect of $\hat{H}_U$ as a $\vex{k}$-independent self-energy correction on the $f$ fermion sector. Then, we show that the quasiparticle dispersion is essentially the same as the noninteracting limit except that the $cf$ hybridization amplitude is renormalized \cite{SM}. Therefore, the low-energy bands are also fragile topological in this perturbative analysis. We suspect the interacting MATBG generally realizes a fragile topological Dirac semimetal at $\nu=0$.

\begin{figure}[t!]
	\includegraphics[width=0.45\textwidth]{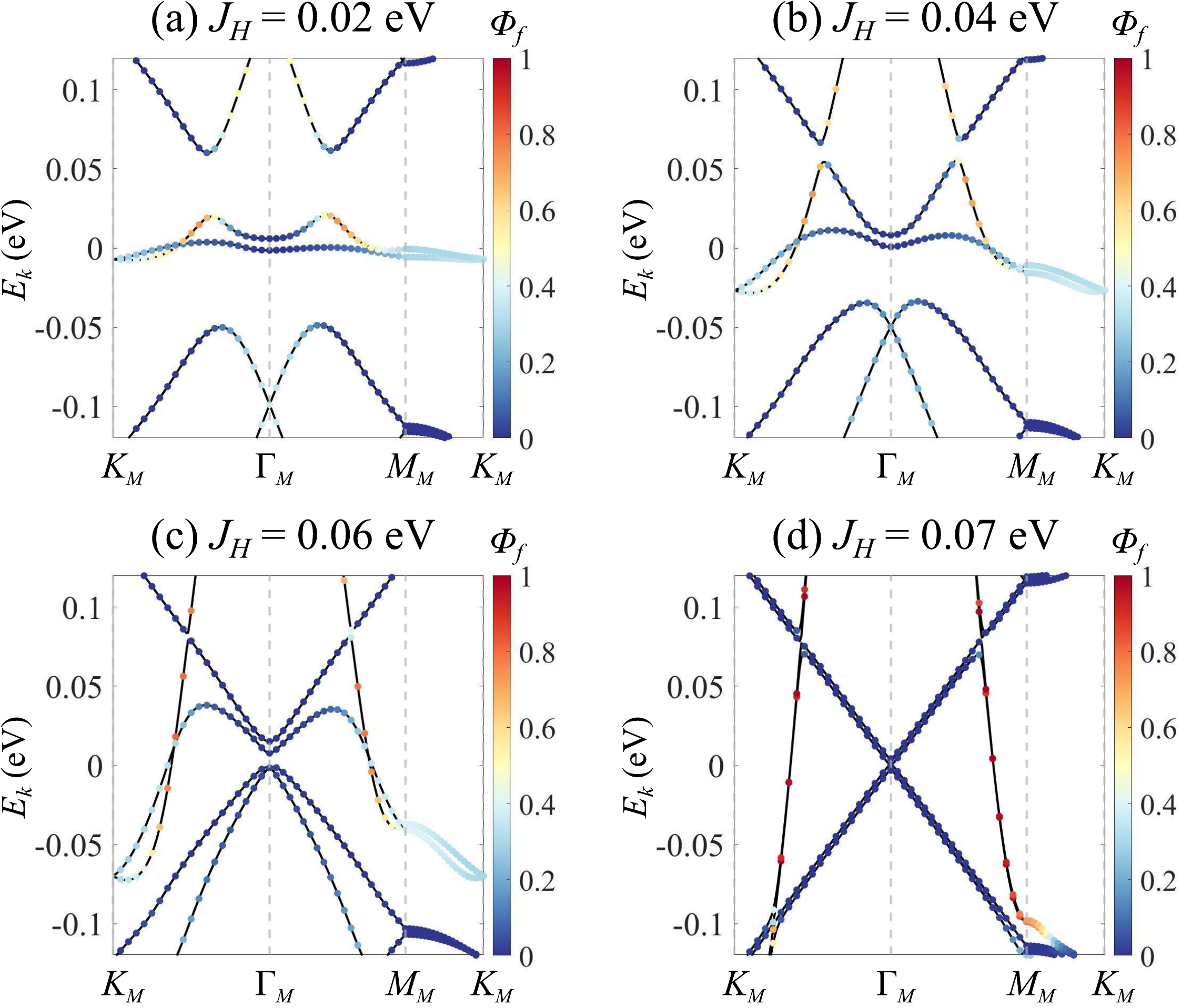}
	\caption{Quasiparticle dispersion of the Kondo-Heisenberg model with $N_f=4$, $J_K=0.03$ eV, and $\eta=+1$. (a) $J_H=0.02$ eV. (b) $J_H=0.04$ eV. (c) $J_H=0.06$ eV. (d) $J_H=0.07$ eV. $\Phi_f$ is the composition of $f$ fermion. $\Phi_f=1$ (red) means the wavefucntions are made of $f$ fermions only; $\Phi_f=0$ (blue) indicates the wavefucntions are made of $c$ fermions only.}
	\label{Fig:Fig3}
\end{figure}	

\textit{Kondo-Heisenberg model.---} One interesting question is if the Kondo semimetal ($\nu=0$) survives perturbations. Particularly, we are interested in the competition between Kondo hybridization and magnetic correlations \cite{Paschen2021quantum}. To this end, we consider $SU(8)$ Heisenberg interaction among the nearest-neighbor sites given by \cite{Coleman2015introduction}
\begin{align}\label{Eq:H_H}
	\hat{H}_H\!=\!J_{H}\!\!\!\sum_{\langle \vex{R},\vex{R}'\rangle}\sum_{\substack{\alpha,\eta,s\\ \alpha'\!,\eta'\!,s'}}\!\!\!\!:f^{\dagger}_{\vex{R},\alpha,\eta,s}f_{\vex{R},\alpha',\eta',s'}::f^{\dagger}_{\vex{R}',\alpha',\eta',s'}f_{\vex{R}',\alpha,\eta,s}:,
\end{align}
where $J_H$ denotes the exchange coupling and $\langle\vex{R},\vex{R}'\rangle$ indicates the nearest-neighbor pairs. Microscopically, the Heisenberg interaction may arise from RKKY and superexchange mechanisms \cite{Coleman2015introduction}. $J_H<0$ corresponds to a ferromagnetic interaction, and we expect that the Heisenberg ineraction with $J_H<0$ drives the system to ground states qualitatively similar to the Hartree-Fock predictions. We thus study on the antiferromagnetic case ($J_H>0$) and focus on the stability of the Kondo semimetal state ($\nu=0$) in the presence of a symmetric magnetic correlation. 
The simplest model is a Kondo-Heisenberg Hamiltonian for MATBG, described by $\hat{H}_{0,c}+\hat{H}_K+\hat{H}_H$ with $J_H>0$. 
To study the antiferromagnetic interaction, we consider a ``bond'' decoupling \cite{Wen2004quantum} as follows:
\begin{align}
	\nonumber&J_H\sum_{\substack{\alpha,\eta,s\\ \alpha'\!,\eta'\!,s'}}\!\!\!\!:f^{\dagger}_{\vex{R},\alpha,\eta,s}f_{\vex{R},\alpha',\eta',s'}::f^{\dagger}_{\vex{R}',\alpha',\eta',s'}f_{\vex{R}',\alpha,\eta,s}:\\
	\label{Eq:H_H_decoupling}\rightarrow&\sum_{\alpha,\eta,s}\left[\chi_{\vex{R},\vex{R}'}f^{\dagger}_{\vex{R}',\alpha,\eta,s}f_{\vex{R},\alpha,\eta,s}+\text{H.c.}\right]+\frac{|\chi_{\vex{R},\vex{R}'}|^2}{J_H},
\end{align} 
where the bond variable
\begin{align}
	\chi_{\vex{R},\vex{R}'}=-J_H\sum_{\alpha,\eta,s}\left\langle f^{\dagger}_{\vex{R},\alpha,\eta,s}f_{\vex{R}',\alpha,\eta,s}\right\rangle.
\end{align}
Our goal is to explore the interplay between Kondo hybridization and magnetic correlation. In particular, we focus only on the situation without symmetry breaking as there is no clear evidence of symmetry breaking at $\nu=0$ of MATBG. As such, we assume uniform and real-valued bond variables \cite{Saremi2007,Hu2022coupled,Coleman2015introduction}, specifically, $\chi_{\vex{R},\vex{R}'}=\chi>0$, corresponding to a formation of a spin liquid with a spinon Fermi surface \cite{Coleman2015introduction,Li2017_Spinon_FS}. We note that the Kondo-Heisenberg model may give rise to a ground state different from the ansatz used here. However, our goal is to explore the interplay between symmetric Kondo hybridization and symmetric magnetic correlation, mainly, how a symmetric magnetic correlation destroys the Kondo semimetal state.

We numerically solve the Kondo-Heisenberg model with the mean-field saddle-point approximation \cite{SM} and compute the hybridization amplitude $|B|$ as well $\chi$. In Fig.~\ref{Fig:Fig2}(a), $|B|$ as a function of $J_H/J_K$ is plotted for a few representative values of $J_K$, showing the breakdown of Kondo hybridization (i.e., $|B|=0$) for a sufficiently large $J_H$. In Fig.~\ref{Fig:Fig2}(b), both $|B|$ and $\chi$ are continuously varying for $J_H\le 0.07$ eV. For $J_H>0.07$ eV, $\chi$ continuously grows as $J_H$ increases, while $|B|$ vanishes to zero. It is also interesting to study how quasiparticle dispersion evolves. In Fig.~\ref{Fig:Fig3}, we show the quasiparticle dispersion along several line cuts ($K_M-\Gamma_M-M_M-K_M$) and use the color to indicate the composition of the $f$ fermion ($\Phi_f$). First, we find that a small $J_H$ [Fig.~\ref{Fig:Fig3}(a)] makes the low-energy bands wider, and the Dirac point is away from the zero energy, suggesting finite Fermi pockets around $K_M$ and $K_M'$ points. As $J_H$ increases, the gaps between low-energy mini bands and the remote bands decrease (as $|B|$ decreases), and the low-energy bandwidth continuously increases (as $\chi$ gives a finite dispersion). In Fig.~\ref{Fig:Fig3}(d), the quasiparticle excitations can be separated by primarily $f$ fermions ($\Phi_f\approx 1$, red dots) and primarily $c$ fermions ($\Phi_f\approx 0$, blue dots), indicative of a negligible $|B|$. Thus, the Fermi surface of the $c$ fermions is reduced to a point at the $\Gamma$ point for a sufficiently large $J_H$.

\textit{Discussion.---} We establish a systematic theory for Kondo lattice models in MATBG, showing a novel route to understand the interaction-driven phenomena in MATBG. Unlike the symmetry-breaking correlated insulators predicted by Hartree-Fock calculations \cite{Xie2020_Correlated_Insulator,Zhang2020_Correlated_ins,Bultinck2020,Liu2021_Th_correlated_ins,TBG_III,TBG_IV,TBG_V,Wagner2022,Song2022,Shi2022HF_Rep}, we find symmetry-preserving correlated metals due to the Kondo coupling between a lattice of $SU(8)$ local moments and delocalized electrons. 
The Kondo lattice description in this Letter may be relevant to the MATBG experiments at $\nu=0$. In particular, the STM measurements concluded the existence of strong correlations at $\nu=0$ \cite{Xie2019_spectroscopic,Wong2020cascade,Choi2019electronic,Jiang2019}, while several transport measurements \cite{Cao2018_tbg1,Cao2018_tbg2,Yankowitz2019,Zondiner2020cascade} did not find a clear sign of symmetry breaking or correlated insulating behavior. Our predicted Dirac Kondo semimetal state potentially explains the dichotomy \cite{Thomson2021} of strong correlation and the symmetry-preserving Dirac point. Our theory also complements the existing list of Kondo lattice systems in the moir\'e materials \cite{Kumar2022,Ramires2021,Guerci2022chiral,Zhao2022gate} and extends the number of topological phases driven by Kondo correlation \cite{Dzero2010,Dzero2016topological,Chang2017mobius,Pixley2017,Lai2018weyl,Paschen2021quantum,Grefe2020,Chen2022topological}. 

One interesting consequence of our results is that the local moments (i.e., $f$ fermions) can be decoupled entirely from the $c$ fermions by a sufficiently strong magnetic correlation, paving the way for realizing a quantum simulator for a $SU(8)$ triangular Heisenberg model in MATBG. The $SU(N)$ magnetic systems with $N>2$ can realize exotic phases, such as spin liquids and valence bond solids \cite{Rokhsar1990,Hermele2009,Gorshkov2010,Hermele2011,Zhang2021,Yao2021}. The ground states of $SU(8)$ triangular Heisenberg model have not been systematically explored (except for $N_f=1$ \cite{Yao2021}), and our work provides a strong incentive for future studies along this direction.

An outstanding question is whether the Kondo lattice model and Kondo-driven correlated states are relevant to the realistic MATBG. Using the parameters in Ref.~\cite{Song2022}, we obtain $U/|\gamma|\approx 2.34$, indicating that the system is in the intermediate coupling regime. Thus, the Kondo lattice models based on the strong-coupling analysis (i.e., $U/|\gamma|\gg 1$) might not explain all the experimental results. Our predicted Dirac Kondo semimetal is consistent with several existing experiments at $\nu=0$, introducing a new perspective to study MATBG \cite{FM_CNP}. On the contrary, at nonzero integer fillings, the Kondo-driven correlated metals cannot explain various symmetry-broken states from experiments. It is possible that other types of interaction-driven states (e.g., correlated insulators) are more energetically favored. Further investigations are required to determine if Kondo-driven correlated states can be realized in the MATBG experiments. Regardless of these complications, our predicted Kondo-driven semimetal provides a plausible novel explanation for semimetallic transport of MATBG at $\nu=0$.

Finally, we discuss a few open questions. In the Kondo-Heisenberg model, we consider a specific magnetic correlation and $J_H>0$. The actual ground state with a finite $J_K$ and $J_H$ should be examined systematically. In particular, for $J_H<0$, we anticipate that the ground states are qualitatively similar to the Hartree-Fock predictions (e.g., valley-polarized states) \cite{Xie2020_Correlated_Insulator,Zhang2020_Correlated_ins,Bultinck2020,Liu2021_Th_correlated_ins,TBG_III,TBG_IV,TBG_V,Wagner2022,Song2022,Shi2022HF_Rep}. It might be interesting to study our Kondo-Heisenberg model with $J_H<0$ for nonzero integer fillings, where symmetry-broken states are reported experimentally. An important task is to compute observables that can reveal the finite-temperature crossover between the coherent Kondo lattice and the decoupled Kondo impurities, such as temperature-dependent tunneling spectroscopy and finite-temperature resistivity.

\begin{acknowledgments}
	\textit{Note added.---} We learned after finishing this work that related results have also recently been obtained by another group (A.M. Tsvelik, Haoyu Hu, and B.A. Bernevig) \cite{Andrei}.
	
	\textit{Acknowledgments.---} Y.-Z.C. thanks Seth Davis, Zhentao Wang, Fengcheng Wu, Ming Xie, Jiabin Yu, and Rui-Xing Zhang for useful discussions.
	This work is supported by the Laboratory for Physical Sciences (Y.-Z.C. and S.D.S.), by JQI-NSF-PFC (Y.-Z.C.), and by ARO W911NF2010232 (Y.-Z.C.). 
\end{acknowledgments}	



	\newpage \clearpage 
	
	\onecolumngrid
	
	\begin{center}
		{\large
			Kondo lattice model in magic-angle twisted bilayer graphene
			\vspace{4pt}
			\\
			SUPPLEMENTAL MATERIAL
		}
	\end{center}
	
	\setcounter{figure}{0}
	\renewcommand{\thefigure}{S\arabic{figure}}
	\setcounter{equation}{0}
	\renewcommand{\theequation}{S\arabic{equation}}	
	
	In this supplemental material, we provide some technical details for the main results in the main text.

\section{Single-particle model}

\begin{figure}[h!]
	\includegraphics[width=0.3\textwidth]{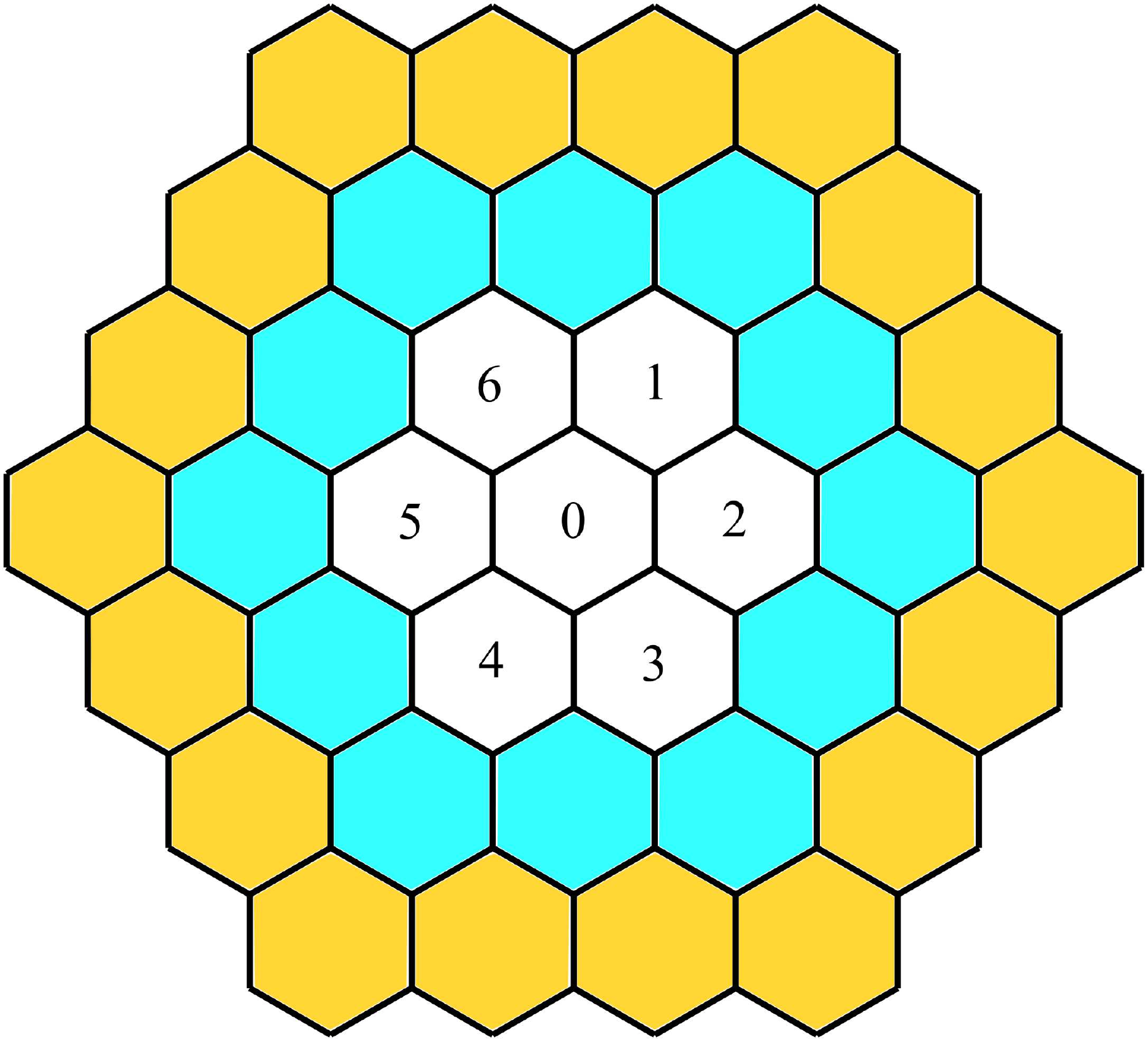}
	\caption{Extended moir\'e Brillouin zones. In the numerical results, a finite number $N_G$ is considered. Including the white regions corresponds to $N_G=7$; including the white and blue regions corresponding to $N_G=19$; including the white, blue, and yellow regions correspond to $N_G=37$.
	}		
	\label{Fig:mBZ}
\end{figure}

The single-particle Hamiltonian in the THF representation \cite{Song2022_S} is given by
\begin{align}
	\nonumber\hat{H}_0=&\sum_{\eta,s,a,a'}\sum_{\vex{q}}\left[h^{(\eta)}_{aa'}(\vex{q})-\mu\delta_{a,a'}\right]c^{\dagger}_{\vex{q},a,\eta,s}c_{\vex{q},a',\eta,s}-\mu\sum_{\alpha,\eta,s}\sum_{\vex{k}\in \text{mBZ}}f^{\dagger}_{\vex{k},\alpha,\eta,s}f_{\vex{k},\alpha,\eta,s}\\
	&+\sum_{\eta,s,\alpha,a}\sum_{\vex{G}}\sum_{\vex{k}\in \text{mBZ}}\!\!\left[V^{(\eta)}_{\alpha a}(\vex{k}\!+\!\vex{G})f^{\dagger}_{\vex{k},\alpha,\eta,s}c_{\vex{k}+\vex{G},a,\eta,s}\!+\!\text{H.c.}
	\right]
\end{align}
where $\mu$ is the chemical potential,
\begin{align}
	\hat{h}^{(\eta)}\!(\vex{q})\!=\!&\left[\!\begin{array}{cc}
		\hat{0}_{2\times 2} & v_*\!\left(\eta q_x\hat{\sigma}_0+iq_y\hat{\sigma}_z\right) \\[2mm]
		v_*\!\left(\eta q_x\hat{\sigma}_0-iq_y\hat{\sigma}_z\right) & M\hat{\sigma}_x		
	\end{array}\!\right],\\
	\hat{V}^{(\eta)}(\vex{q})=&e^{-\frac{|\vex{q}|^2\lambda^2}{2}}\!\!\left[\!\begin{array}{cc}
		\gamma\hat{\sigma}_0 + v_*'\!\left(\eta q_x\hat{\sigma}_x+q_y\hat{\sigma}_y\right), & \hat{0}_{2\times2}
	\end{array}
	\!\right].
\end{align}
In the above expressions, $\hat{\sigma}_{0}$ and $\hat{\sigma}_{\mu}$ represent the $2\times2$ identity operator and $\mu$-component of Pauli matrix respectively, $v_*=-4.303$ eV \AA, $M=3.697$ meV, $\gamma=-24.75$ meV, $v_*'=1.622$ eV \AA, and $\lambda=0.3375a_M$ \cite{Song2022_S}. The parameters correspond to $w_0/w_1=0.8$, $w_1=110$ meV, and $\theta=1.05^{\circ}$ in the BM model \cite{Song2022_S}. In our calculations, we consider a finite number ($N_G$) of mBZs. Intuitively, one should choose $N_G=7,19,37,\dots$ as shown in Fig.~\ref{Fig:mBZ}.
We find that $N_G\ge 7$ is required for recovering the Dirac points in the mini bands of MATBG, and $N_G=37$ is used in all the results presented in the main text.

The single-particle spectrum can be obtained by solving the characteristic equation (with $\mu=0$) as follows:
\begin{align}
	\det\left(\omega\hat{1}_{(2+4N_G)\times(2+4N_G)}-\left[\begin{array}{cc}
		\hat{0}_{2\times 2} & \hat{\mathcal{V}}^{(\eta)}(\vex{k})\\
		\left[\hat{\mathcal{V}}^{\eta}(\vex{k})\right]^{\dagger} & \hat{\mathcal{H}}^{(\eta)}_{c}(\vex{k})
	\end{array}\right]\right)=0,
\end{align}
where
\begin{align}
	\hat{\mathcal{V}}^{(\eta)}=&\left[\begin{array}{cccc}
		\hat{V}^{(\eta)}(\vex{k}+\vex{G}_0) & \hat{V}^{(\eta)}(\vex{k}+\vex{G}_1) & \dots & \hat{V}^{(\eta)}(\vex{k}+\vex{G}_{N_G-1})
	\end{array}
	\right],\\
	\hat{\mathcal{H}}^{(\eta)}_c(\vex{k})=&\left[\begin{array}{cccc}
		\hat{h}^{(\eta)}(\vex{k}+\vex{G}_0) & 0 & \dots & 0\\
		0 & \hat{h}^{(\eta)}(\vex{k}+\vex{G}_1) &\dots & 0\\
		\vdots & \vdots & \ddots & \vdots\\
		0 & 0 & \dots & \hat{h}^{(\eta)}(\vex{k}+\vex{G}_{N_G-1})
	\end{array}
	\right].
\end{align}

Finally, the Fourier transform convention for the $f$ fermions is as follows:
\begin{align}
	f_{\vex{R},\alpha,\eta,s}=\frac{1}{\sqrt{\mathcal{N}_k}}\sum_{\vex{k}\in \text{mBZ}}e^{i\vex{k}\cdot\vex{R}}f_{\vex{k},\alpha,\eta,s},\,\,\,f_{\vex{k},\alpha,\eta,s}=\frac{1}{\sqrt{\mathcal{N}_k}}\sum_{\vex{R}}e^{-i\vex{k}\cdot\vex{R}}f_{\vex{R},\alpha,\eta,s},
\end{align}
where $\mathcal{N}_k$ is the number of $\vex{k}$ points in the first mBZ, equivalent to the number of superlattice points.

\section{Coulomb interaction}

We consider the Coulomb interaction given by
\begin{align}
	\hat{H}_I=\frac{1}{2}\int\limits_{\vex{r},\vex{r}'}:\hat{\rho}_0(\vex{r}):V(\vex{r}_1-\vex{r}_2):\hat{\rho}_0(\vex{r}'):,
\end{align}
where $\hat{\rho}_0(\vex{r})$ is the long-wavelength component of density operator incorporating microscopic labels $l$, $\beta$, $\eta$, $s$ and $:\hat{\rho}_0(\vex{r}):\equiv \hat{\rho}_0(\vex{r}) -\langle G|\hat{\rho}_0(\vex{r})|G\rangle$ with $|G\rangle$ being the normal state at charge neutrality. The subtraction of normal state at charge neutrality is important as the ionic background also contributes to the Coulomb potential. The expectation values are given by
\begin{align}
	\langle G|f^{\dagger}_{\vex{R},\alpha,\eta,s}f_{\vex{R}',\alpha',\eta',s'}|G\rangle=&\frac{1}{2}\delta_{\vex{R},\vex{R}'}\delta_{\alpha,\alpha'}\delta_{\eta,\eta'}\delta_{s,s'},\\
	\langle G|c^{\dagger}_{\vex{k},a,\eta,s}c_{\vex{k}',a',\eta',s'}|G\rangle=&\frac{1}{2}\delta_{\vex{k},\vex{k}'}\delta_{a,a'}\delta_{\eta,\eta'}\delta_{s,s'},\\
	\langle G|f^{\dagger}_{\vex{R},\alpha,\eta,s}c_{\vex{k}',a',\eta',s'}|G\rangle=&\langle G|c_{\vex{k}',\vex{Q}',a',\eta',s'}f_{\vex{R},\alpha,\eta,s}|G\rangle=0
\end{align}

In Ref.~\cite{Song2022_S}, the Coulomb interaction can be projected into the THF basis, and $\hat{H}_I=\hat{H}_U+\hat{H}_V+\hat{H}_W+\hat{H}_J$, where
\begin{align}
	\hat{H}_U=&\frac{U}{2}\sum_{\vex{R}}:\hat{\rho}_{\vex{R}}^f::\hat{\rho}_{\vex{R}}^f:+\frac{U_2}{2}\sum_{\left\langle\vex{R},\vex{R}'\right\rangle}:\hat{\rho}_{\vex{R}}^f::\hat{\rho}_{\vex{R}'}^f:,\\
	\hat{H}_V=&\frac{1}{2}\int\limits_{\vex{r},\vex{r}'}:\hat{\rho}^c(\vex{r}):V(\vex{r}-\vex{r}'):\hat{\rho}^c(\vex{r}'):,\\
	\hat{H}_W=&\Omega_0\sum_{\vex{R},a}W_a:\hat{\rho}_{\vex{R}}^f::\hat{\rho}_{a}^c(\vex{R}):,\\
	\label{Eq:H_J_S}\hat{H}_J
	=&-\frac{J}{\mathcal{N}_k}\sum_{\eta,\alpha,s_1,s_2}\sum_{\vex{G},\vex{G}'}\,\sum_{\vex{k},\vex{k}'\in \text{mBZ}}e^{-i(\vex{k}'-\vex{k})\cdot\vex{R}}\left[
	\begin{array}{c}
		:f^{\dagger}_{\vex{R},\alpha,\eta,s_1}f_{\vex{R},\alpha,\eta,s_2}::c^{\dagger}_{\vex{k}'+\vex{G}',\alpha+2,\eta,s_2}c_{\vex{k}+\vex{G},\alpha+2,\eta,s_1}:\\[2mm]
		-f^{\dagger}_{\vex{R},\bar{\alpha},-\eta,s_1}f_{\vex{R},\alpha,\eta,s_2}c^{\dagger}_{\vex{k}'+\vex{G}',\alpha+2,\eta,s_2}c_{\vex{k}+\vex{G},\bar{\alpha}+2,-\eta,s_1}
	\end{array}
	\right].
\end{align}
In the above expressions, $\Omega_0$ is the area of a moir\'e unit cell, $\hat{\rho}^f_{\vex{R}}$ denotes the local density of $f$ fermions, $\hat{\rho}^c(\vex{r})$ denotes the local density of $c$ fermions, $\bar{1}=2$, and $\bar{2}=1$.
In Ref.~\cite{Song2022_S}, the coupling constants are estimated as follows: $U=57.95$ meV, $U_2=2.33$ meV, $W_{1,2}=44.03$ meV, $W_{3,4}=50.2$ meV, and $J=16.38$ meV. Since $U_2\ll U$, we ignore the $U_2$ term completely. We also note that the values of interaction terms are extracted with $w_0/w_1=0.8$, $w_1=110$ meV, and $\theta=1.05^{\circ}$ in the BM model \cite{Song2022_S}. As pointed out in \cite{Song2022_S}, the chiral limit ($w_0$) gives a much less delocalized $f$ orbital such that $U\ll\gamma$. We focus only on the situation relevant to the MATBG experiments.

\section{Weak coupling analysis}

Following Refs. \cite{Edwards1988transition_S,Hewson1997kondo_S}, we generalize the formalism for the single-band periodic Anderson model to $\hat{H}_{0,c}+\hat{H}_{0,cf}+\hat{H}_U$ in this section. The main idea is that a self-energy correction can be obtained by perturbing the onsite Hubbard interaction, and then we study the general properties of the quasiparticle dispersion. The retarded one-electron Green function can be determined by
\begin{align}\label{Eq:GF_Edwards}
	\left[\begin{array}{cc}
		\omega-\hat{\Sigma}^{(\eta)}_s(\omega,\vex{k}) & -\hat{\mathcal{V}}^{(\eta)}(\vex{k}) \\[1mm]
		-\left[\hat{\mathcal{V}}^{(\eta)}(\vex{k})\right]^{\dagger} & \omega-\hat{\mathcal{H}}^{(\eta)}_c(\vex{k})
	\end{array}
	\right]\left[\begin{array}{cc}
		\hat{G}^{(ff,\eta)}_s(\omega,\vex{k}) & \hat{G}^{(fc,\eta)}_s(\omega,\vex{k})\\[1mm]
		\hat{G}^{(cf,\eta)}_s(\omega,\vex{k}) & \hat{G}^{(cc,\eta)}_s(\omega,\vex{k})
	\end{array}\right]=\hat{1}_{(2+4N_G)\times (2+4N_G)},
\end{align}
where $\hat{\Sigma}^{(\eta)}_s$ is the proper self energy for valley $\eta$ and spin $s$, $\hat{G}_s^{(ff,\eta)}$, $\hat{G}_s^{(fc,\eta)}$, $\hat{G}_s^{(cf,\eta)}$, and $\hat{G}_s^{(cc,\eta)}$ are the retarded Green functions. According to Refs. \cite{Edwards1988transition_S,Hewson1997kondo_S}, equation~(\ref{Eq:GF_Edwards}) represents the minimal formalism that realizes the heavy fermion physics. The $\hat{\Sigma}^{(\eta)}_s$ can be in principle derived by treating $\hat{H}_U$ perturbatively, and $\hat{\Sigma}_s(\omega,\vex{k})$ vanishes when $U=0$. 
The self energy $\hat{\Sigma}_s^{(\eta)}$ is diagonal in the orbital space as there is no nontrivial dynamics between two different orbitals. Thus, $\hat{\Sigma}_{s,\alpha \alpha'}^{(\eta)}=\Sigma^{(\eta)}_{s,\alpha\alpha'}\delta_{\alpha, \alpha'}$. Assuming that $\Sigma^{(\eta)}_{s,\alpha \alpha'}$ can be expanded in a Taylor's series about $\omega=\mathcal{E}_F$ and $\vex{k}=\vex{k}_F$, we obtain (with $\eta$ superscript dropped)
\begin{align}
	\Sigma_{s,\alpha \alpha}(\omega,\vex{k})\approx\Sigma^R_{s,\alpha \alpha}(\mathcal{E}_F,\vex{k}_F)+\left(\vex{k}-\vex{k}_F\right)\cdot\vex{\nabla}\Sigma^R_{s,\alpha \alpha}(\mathcal{E}_F,\vex{k})\Big|_{\vex{k}=\vex{k}_F}+\left(\omega-\mathcal{E}_F\right)\partial_{\omega}\Sigma^R_{s,\alpha \alpha}(\omega,\vex{k}_F)\Big|_{\omega=\mathcal{E}_F}+\dots,
\end{align}
where $\Sigma_{s,\alpha \alpha}^R$ denotes the real part of $\Sigma_{s,\alpha \alpha}$. Based on Luttinger's results \cite{Hewson1997kondo_S}, the imaginary part of self energy $\Sigma^I_s(\omega,\vex{k}_F)\sim\left(\omega-\mathcal{E}_F\right)^2$ quite generally. If we retain only terms to first order in $\omega-\mathcal{E}_F$ and $|\vex{k}-\vex{k}_F|$, 
\begin{align}
	\nonumber&\omega-\hat{\Sigma}_{s,\alpha\alpha}^{(\eta)}(\omega,\vex{k})\\
	\nonumber\approx&\omega -\Sigma^R_{s,\alpha\alpha}(\mathcal{E}_F,\vex{k}_F)-\left(\vex{k}-\vex{k}_F\right)\cdot\vex{\nabla}\Sigma^R_{s,\alpha\alpha}(\mathcal{E}_F,\vex{k})\Big|_{\vex{k}=\vex{k}_F}-\left(\omega-\mathcal{E}_F\right)\partial_{\omega}\Sigma^R_{s,\alpha\alpha}(\omega,\vex{k}_F)\Big|_{\omega=\mathcal{E}_F}\\
	\nonumber=&\left[1-\partial_{\omega}\Sigma^R_{s,\alpha\alpha}(\omega,\vex{k}_F)\Big|_{\omega=\mathcal{E}_F}\right]\omega+\mathcal{E}_F\partial_{\omega}\Sigma^R_{s,\alpha\alpha}(\omega,\vex{k}_F)\Big|_{\omega=\mathcal{E}_F}-\Sigma^R_{s,\alpha\alpha}(\mathcal{E}_F,\vex{k}_F)-\left(\vex{k}-\vex{k}_F\right)\cdot\vex{\nabla}\Sigma^R_{s,\alpha\alpha}(\mathcal{E}_F,\vex{k})\Big|_{\vex{k}=\vex{k}_F}\\
	\equiv&Z_{\vex{k}_F}^{-1}\left[\omega-\mathcal{E}_F-\tilde{\epsilon}_{f}(\vex{k})\right],
\end{align}
where
\begin{align}
	Z_{\vex{k}_F}=&\left[1-\partial_{\omega}\Sigma^R_{s,\alpha\alpha}(\omega,\vex{k}_F)\Big|_{\omega=\mathcal{E}_F}\right]^{-1},\\
	\tilde{\epsilon}_{f}(\vex{k})=&Z_{\vex{k}_F}\left[\Sigma^R_{s,\alpha\alpha}(\mathcal{E}_F,\vex{k}_F)+\left(\vex{k}-\vex{k}_F\right)\cdot\vex{\nabla}\Sigma^R_{s,aa}(\mathcal{E}_F,\vex{k})\Big|_{\vex{k}=\vex{k}_F}-\mathcal{E}_F\right].
\end{align}
We assume that $\Sigma^{(\eta)}_{s,11}=\Sigma^{(\eta)}_{s,22}$ as there is no clear distinction between two orbitals.

One of the important results based on this formalism is the quasiparticle dispersion, corresponding to the characteristic equation as follows:
\begin{align}
	&\det\left(\left[\begin{array}{cc}
		Z^{-1}_{\vex{k}_F}\left[\omega-\mathcal{E}_F-\tilde{\epsilon}_f(\vex{k})\right]\hat{1}_{2\times 2} & -\hat{\mathcal{V}}^{(\eta)}(\vex{k})\\
		-\left[\hat{\mathcal{V}}^{\eta}(\vex{k})\right]^{\dagger} &\omega\hat{1}_{4N_G\times 4NG}- \hat{\mathcal{H}}^{(\eta)}_{c}(\vex{k})
	\end{array}\right]\right)=0\\
	\rightarrow&\det\left(Z^{-1}_{\vex{k}_F}\left[\omega-\mathcal{E}_F-\tilde{\epsilon}_f(\vex{k})\right]\hat{1}_{2\times 2}\right)
	\det\left(\omega\hat{1}_{4N_G\times 4NG}-\hat{\mathcal{H}}^{(\eta)}_{c}(\vex{k})-\frac{Z_{\vex{k}_F}}{\omega-\mathcal{E}_F-\tilde{\epsilon}_f(\vex{k})}\left[\hat{\mathcal{V}}^{(\eta)}(\vex{k})\right]^{\dagger}\hat{\mathcal{V}}^{(\eta)}(\vex{k})
	\right)=0.
\end{align}
The above expression is very similar to the characteristic equation for the single-particle case.
In particular, if $\mathcal{E}_F=0$ and $\tilde{\epsilon}_f(\vex{k})= 0$ (corresponding to $\nu=0$ of MATBG), the quasiparticle band is almost identical to the single particle band structure except that the $\hat{\mathcal{V}}^{(\eta)}$ is replaced by $\sqrt{Z_{\vex{k}_F}}\hat{\mathcal{V}}^{(\eta)}$. Thus, the quasiparticle dispersion here is a Dirac semimetal with the fragile topology. We note that the mean-field approach also gives rise to a fragile topological Dirac semimetal dispersion at $\nu=0$.

\section{Derivation of Kondo coupling}

When the number of $f$ fermion is frozen at an impurity site, the Anderson impurity model can be effectively described by a Kondo model. We provide a detailed derivation in this section. First, the energy of the impurity site is given by $E(N_f)=-N_f\mu+\frac{U}{2}(N_f-4)^2$. $\mu$ here is the local chemical potential for the impurity site which can be very different from the chemical potential for the delocalized $c$ fermions. We inspect the interaction energy differences between different number of $f$ fermions.
\begin{align}
	E(N_f-1)-E(N_f)=&\left[\frac{U}{2}\left(N_f-1-4\right)^2-(N_f-1)\mu\right]-\left[\frac{U}{2}\left(N_f-4\right)^2-N_f\mu\right]=U\left(\frac{9}{2}-N_f\right)+\mu,\\
	E(N_f+1)-E(N_f)=&\left[\frac{U}{2}\left(N_f+1-4\right)^2-(N_f+1)\mu\right]-\left[\frac{U}{2}\left(N_f-4\right)^2-N_f\mu\right]
	=U\left(N_f-\frac{7}{2}\right)-\mu.
\end{align}
The conditions for exactly $N_f$ in an impurity site correspond to $E_I(N_f-1)-E_I(N_f)>0$ and $E(N_f+1)-E(N_f)>0$. Thus, 
\begin{align}\label{Eq:criteria_Kondo}
	N_f-\frac{9}{2}<\mu/U<N_f-\frac{7}{2}.
\end{align}
For example, $N_f=4$ corresponds to $|\mu|<U/2$; $N_f=3$ corresponds to $-3U/2<\mu<-U/2$; $N_f=2$ corresponds to $-5U/2<\mu<-3U/2$;  $N_f=1$ corresponds to $-7U/2<\mu<-5U/2$. Without loss of generality, we can choose $\mu=(N_f-4)U$ for all cases.

To derive the Kondo coupling, we assume $N_f$ fermions in the impurity site. Then, the Kondo coupling can be derived via Schrieffer-Wolf transformation (or second-order perturbation theory) as follows:
\begin{align}
	\hat{H}_{\text{K}}=&-\frac{1}{\mathcal{N}_k}\sum_{\vex{R}}\sum_{\substack{\alpha,\eta,a,s\\ \alpha',\eta',a',s'}}\sum_{\vex{G},\vex{G}'}\sum_{\vex{k},\vex{k}'}e^{i(\vex{k}-\vex{k}')\cdot\vex{R}}\left[V^{(\eta')}_{\alpha'a'}(\vex{k}'\!+\!\vex{G}')\right]^*V^{(\eta)}_{\alpha a}(\vex{k}\!+\!\vex{G})
	\left[\begin{array}{r}
		\frac{\left(c^{\dagger}_{\vex{k}'+\vex{G}',a',\eta',s'}f_{\vex{R},\alpha',\eta',s'}\right)\left(f^{\dagger}_{\vex{R},\alpha,\eta,s}c_{\vex{k}+\vex{G},a,\eta,s}\right)}{U\left(N_f-\frac{7}{2}\right)-\mu}\\[2mm]
		+\frac{\left(f^{\dagger}_{\vex{R},\alpha,\eta,s}c_{\vex{k}+\vex{G},a,\eta,s}\right)\left(c^{\dagger}_{\vex{k}'+\vex{G}',a',\eta',s'}f_{\vex{R},\alpha',\eta',s'}\right)}{U\left(\frac{9}{2}-N_f\right)+\mu}
	\end{array}
	\right]\hat{\mathcal{P}}_{N_f}\\
	\label{Eq:H_K_S}=&\frac{1}{\mathcal{N}}\sum_{\vex{R}}\sum_{\substack{\alpha,\eta,a,s\\ \alpha',\eta',a',s'}}\sum_{\vex{G},\vex{G}'}\sum_{\vex{k},\vex{k}'}\frac{U \left[V^{(\eta')}_{\alpha'a'}(\vex{k}'\!+\!\vex{G}')\right]^*V^{(\eta)}_{\alpha a}(\vex{k}\!+\!\vex{G})}{\left[U\left(\frac{9}{2}-N_f\right)+\mu\right]\left[U\left(N_f-\frac{7}{2}\right)-\mu\right]} e^{i(\vex{k}-\vex{k}')\cdot\vex{R}}:f^{\dagger}_{\vex{R},\alpha,\eta,s}f_{\vex{R},\alpha',\eta',s'}::c^{\dagger}_{\vex{k}'+\vex{G}',a',\eta',s'}c_{\vex{k}+\vex{G},a,\eta,s}:\hat{\mathcal{P}}_{N_f}
\end{align}
where $\hat{\mathcal{P}}_{N_f}$ is the projection operator onto the subspace with exactly $N_f$ localized $f$ fermions. 
When $\vex{q},\vex{q}'\rightarrow 0$, the coupling constant is reduced to
\begin{align}\label{Eq:Kondo_coupling}
	\frac{U \left[V^{(\eta')}_{\alpha'a'}(\vex{q}')\right]^*V^{(\eta)}_{\alpha a}(\vex{q})}{\left[U\left(\frac{9}{2}-N_f\right)+\mu\right]\left[U\left(N_f-\frac{7}{2}\right)-\mu\right]}\Bigg|_{\vex{q},\vex{q}'\rightarrow 0}=\frac{U\gamma^2}{\left[U\left(\frac{9}{2}-N_f\right)+\mu\right]\left[U\left(N_f-\frac{7}{2}\right)-\mu\right]}\delta_{a\alpha}\delta_{\alpha'a'}\equiv J_K\delta_{a\alpha}\delta_{\alpha'a'}.
\end{align}
Using $\mu=(N_f-4)U$, we find that $J_K=4\gamma^2/U$ independent of $N_f$.
Equation (\ref{Eq:H_K_S}) is a $SU(8)$ Coqblin-Schrieffer coupling. To derive Eq.~(\ref{Eq:H_K_S}), we have used the Fierz identity for $SU(n)$ group: 
\begin{align}
	\sum_{a=1}^{n^2-1}t_{ij}^at_{kl}^a=\frac{1}{2}\left(\delta_{il}\delta_{jk}-\frac{1}{n}\delta_{ij}\delta_{kl}\right),
\end{align}
where $t_{ij}^a$ is the Lie algebra generators in the fundamental representation satisfying $\text{Tr}\left(t^at^b\right)=\frac{1}{2}\delta^{ab}$. With the Fiertz identity, we obtain
\begin{align}
	f^{\dagger}_{\vex{R},i}f_{\vex{R},j}c^{\dagger}_{\vex{k}',k}c_{\vex{k},l}\delta_{i,l}\delta_{j,k}=&2\sum_bf^{\dagger}_{\vex{R},i}t_{ij}^bf_{\vex{R},j}c^{\dagger}_{\vex{k}',k}t_{kl}^bc_{\vex{k},l}+\frac{1}{8}f^{\dagger}_{\vex{R},i}f_{\vex{R},j}c^{\dagger}_{\vex{k}',k}c_{\vex{k},l}\delta_{i,j}\delta_{k,l}\\
	=&2\sum_bf^{\dagger}_{\vex{R},i}t_{ij}^bf_{\vex{R},j}c^{\dagger}_{\vex{k}',k}t_{kl}^bc_{\vex{k},l}+\frac{N_f}{8}c^{\dagger}_{\vex{k}',l}c_{\vex{k},l},
\end{align}
where we have used $i,j,k,l$ indices as shorthand notation of $a,\eta,s$ and summed repeated indices. 

\section{Absence of $\hat{H}_J$ contribution in the hybridization decoupling}

In Ref.~\cite{Song2022_S}, the $\hat{H}_J$ term plays a crucial role in the Hartree-Fock calculations. In this section, we explain why $\hat{H}_J$ is irrelevant to the Kondo lattice problem without any symmetry breaking. The $\hat{H}_J$ term [Eq.~(\ref{Eq:H_J_S})] is given by \cite{Song2022_S}:
\begin{align}
	\hat{H}_J=&-\frac{J}{\mathcal{N}_k}\sum_{\eta,\alpha,s_1,s_2}\sum_{\vex{G},\vex{G}'}\,\sum_{\vex{k},\vex{k}'\in \text{mBZ}}e^{-i(\vex{k}'-\vex{k})\cdot\vex{R}}\left[
	\begin{array}{c}
		:f^{\dagger}_{\vex{R},\alpha,\eta,s_1}f_{\vex{R},\alpha,\eta,s_2}::c^{\dagger}_{\vex{k}'+\vex{G}',\alpha+2,\eta,s_2}c_{\vex{k}+\vex{G},\alpha+2,\eta,s_1}:\\[2mm]
		-f^{\dagger}_{\vex{R},\bar{\alpha},-\eta,s_1}f_{\vex{R},\alpha,\eta,s_2}c^{\dagger}_{\vex{k}'+\vex{G}',\alpha+2,\eta,s_2}c_{\vex{k}+\vex{G},\bar{\alpha}+2,-\eta,s_1}
	\end{array}
	\right]\\
	=&+\frac{J}{\mathcal{N}_k}\sum_{\eta,\alpha,s_1,s_2}\sum_{\vex{G},\vex{G}'}\,\sum_{\vex{k},\vex{k}'\in \text{mBZ}}e^{-i(\vex{k}'-\vex{k})\cdot\vex{R}}\left[
	\begin{array}{c}
		\frac{1}{2}f^{\dagger}_{\vex{R},\alpha,\eta,s_1}c_{\vex{k}+\vex{G},\alpha+2,\eta,s_1}c^{\dagger}_{\vex{k}'+\vex{G}',\alpha+2,\eta,s_2}f_{\vex{R},\alpha,\eta,s_2}\\[2mm]
		+\frac{1}{2}c^{\dagger}_{\vex{k}'+\vex{G}',\alpha+2,\eta,s_2}f_{\vex{R},\alpha,\eta,s_2}f^{\dagger}_{\vex{R},\alpha,\eta,s_1}c_{\vex{k}+\vex{G},\alpha+2,\eta,s_1}\\[2mm]
		-f^{\dagger}_{\vex{R},\bar{\alpha},-\eta,s_1}c_{\vex{k}+\vex{G},\bar{\alpha}+2,-\eta,s_1}c^{\dagger}_{\vex{k}'+\vex{G}',\alpha+2,\eta,s_2}f_{\vex{R},\alpha,\eta,s_2}
	\end{array}
	\right].
\end{align}
We can treat the $\hat{H}_J$ term as two separate interactions -- A ferromagnetic coupling and an antiferromagnetic coupling. Now, we employ the mean-field hybridizaton decoupling. Using $AB\approx \langle A\rangle B+A\langle B\rangle-\langle A\rangle\langle B\rangle$, the $\hat{H}_J$ term can be approximated by
\begin{align}\label{Eq:H_J_MF}
	\hat{H}_J\rightarrow \frac{J}{\mathcal{N}_k}\sum_{\eta,\alpha,s_1,s_2}\sum_{\vex{G},\vex{G}'}\,\sum_{\vex{k},\vex{k}'\in \text{mBZ}}e^{-i(\vex{k}'-\vex{k})\cdot\vex{R}}\left[
	\begin{array}{c}
		\langle f^{\dagger}_{\vex{R},\alpha,\eta,s_1}c_{\vex{k}+\vex{G},\alpha+2,\eta,s_1}\rangle c^{\dagger}_{\vex{k}'+\vex{G}',\alpha+2,\eta,s_2}f_{\vex{R},\alpha,\eta,s_2}\\[2mm]
		+f^{\dagger}_{\vex{R},\alpha,\eta,s_1}c_{\vex{k}+\vex{G},\alpha+2,\eta,s_1}\langle c^{\dagger}_{\vex{k}'+\vex{G}',\alpha+2,\eta,s_2}f_{\vex{R},\alpha,\eta,s_2}\rangle\\[2mm]
		-\langle f^{\dagger}_{\vex{R},\alpha,\eta,s_1}c_{\vex{k}+\vex{G},\alpha+2,\eta,s_1}\rangle \langle c^{\dagger}_{\vex{k}'+\vex{G}',\alpha+2,\eta,s_2}f_{\vex{R},\alpha,\eta,s_2}\rangle\\[2mm]
		-\langle f^{\dagger}_{\vex{R},\bar{\alpha},-\eta,s_1}c_{\vex{k}+\vex{G},\bar{\alpha}+2,-\eta,s_1}\rangle c^{\dagger}_{\vex{k}'+\vex{G}',\alpha+2,\eta,s_2}f_{\vex{R},\alpha,\eta,s_2}\\[2mm]
		-f^{\dagger}_{\vex{R},\bar{\alpha},-\eta,s_1}c_{\vex{k}+\vex{G},\bar{\alpha}+2,-\eta,s_1}\langle c^{\dagger}_{\vex{k}'+\vex{G}',\alpha+2,\eta,s_2}f_{\vex{R},\alpha,\eta,s_2}\rangle\\[2mm]
		+\langle f^{\dagger}_{\vex{R},\bar{\alpha},-\eta,s_1}c_{\vex{k}+\vex{G},\bar{\alpha}+2,-\eta,s_1}\rangle\langle c^{\dagger}_{\vex{k}'+\vex{G}',\alpha+2,\eta,s_2}f_{\vex{R},\alpha,\eta,s_2}\rangle
	\end{array}
	\right].
\end{align}
For symmetric Kondo-driven states, $\langle f^{\dagger}_{\vex{R},\alpha,\eta,s}c_{\vex{k}+\vex{G},\alpha+2,\eta,s}\rangle$ is independent of $\alpha$, $\eta$, or $s$, resulting in a vanishing Eq.~(\ref{Eq:H_J_MF}). Thus, we ignore $\hat{H}_J$ at all in the main text because we focus only on the symmetry-preserving Kondo-driven correlated states.

\section{Imaginary-time path integral and mean-field approximation}

In this section, we discuss how to study the Kondo-Heisenberg model given by $\hat{H}_{0,c}+\hat{H}_{K}+\hat{H}_H$. We construct a path integral formalsim incorporating the local constraints.
In the imaginary-time path integral, 
\begin{align}
	Z=&\int\mathcal{D}\left[c,\bar{c},f,\bar{f};\mu_f,\mu_c\right]e^{-\mathcal{S}},\\
	\nonumber\mathcal{S}=&\int\limits_{\tau}\sum_{\eta,s}\sum_{a,a'}\sum_{\vex{q}}\bar{c}_{\vex{q},a,\eta,s}\left[\partial_{\tau}+\hat{h}^{(\eta)}_{a,a'}(\vex{q})-\mu_c\right]c_{\vex{q},a',\eta,s}+\int\limits_{\tau}\sum_{\vex{R}}\sum_{\alpha,\eta,s}\bar{f}_{\vex{R},\alpha,\eta,s}\partial_{\tau}f_{\vex{R},\alpha,\eta,s}\\
	\nonumber&-\frac{J_K}{\mathcal{N}_k}\!\!\int\limits_{\tau}\!\!\sum_{\vex{R}}\sum_{\vex{G},\vex{G}'}\sum_{\vex{k},\vex{k}'\in\text{mBZ}}\sum_{\substack{\alpha,\eta,a,s\\ \alpha',\eta',a',s'}}\frac{\left[V^{(\eta')}_{\alpha'a'}(\vex{k}'\!+\!\vex{G}')\right]^*V^{(\eta)}_{\alpha a}(\vex{k}\!+\!\vex{G})}{\gamma^2}e^{i\left(\vex{k}-\vex{k}'\right)\cdot\vex{R}}\bar{f}_{\vex{R},\alpha,\eta,s}c_{\vex{k}+\vex{G},a,\eta,s}\bar{c}_{\vex{k}'+\vex{G}',a',\eta',s'}f_{\vex{R},\alpha',\eta',s'}\\
	\nonumber&-J_H\int\limits_{\tau}\sum_{\langle \vex{R},\vex{R}'\rangle}\sum_{\substack{\alpha,\eta,s\\ \alpha',\eta',s'}}\bar{f}_{\vex{R},\alpha,\eta,s}f_{\vex{R}',\alpha,\eta,s}\bar{f}_{\vex{R}',\alpha',\eta',s'}f_{\vex{R},\alpha',\eta',s'}\\
	&-\int\limits_{\tau}\sum_{\vex{R}}\mu_f(\vex{R})\left(\sum_{\alpha,\eta,s}\bar{f}_{\vex{R},\alpha,\eta,s}f_{\vex{R},\alpha,\eta,s}-N_f\right)+\beta\mu_c\mathcal{N}_kN_c,
\end{align}
where $\mu_f(\vex{R})$ and $\mu_c$ are the Lagrangian multipliers ensuring the numbers of $f$ and $c$ fermions. $\nu_f=N_f-4$ and $\nu_c=N_c-8N_G$. We note that $\mu_f(\vex{R})$ can be very different from the $\mu$ in Eq.~(\ref{Eq:criteria_Kondo}) because of the strong renormalization effect. 

Using Hubbard-Stratonovich decoupling, we replace the $J_K$ and $J_H$ terms by
\begin{align}
	\nonumber
	&-\frac{J_K}{\mathcal{N}_k}\!\!\int\limits_{\tau}\!\!\sum_{\vex{R}}\sum_{\vex{G},\vex{G}'}\sum_{\vex{k},\vex{k}'\in\text{mBZ}}\sum_{\substack{\alpha,\eta,a,s\\ \alpha',\eta',a',s'}}\frac{\left[V^{(\eta')}_{\alpha'a'}(\vex{k}'\!+\!\vex{G}')\right]^*V^{(\eta)}_{\alpha a}(\vex{k}\!+\!\vex{G})}{\gamma^2}e^{i\left(\vex{k}-\vex{k}'\right)\cdot\vex{R}}\bar{f}_{\vex{R},\alpha,\eta,s}c_{\vex{k}+\vex{G},a,\eta,s}\bar{c}_{\vex{k}'+\vex{G}',a',\eta',s'}f_{\vex{R},\alpha',\eta',s'}\\
	\rightarrow&\!\int\limits_{\tau}\!\sum_{\vex{R}}\!\left\{\!\!\frac{1}{\sqrt{\mathcal{N}_k}}\sum_{\vex{G}}\sum_{\vex{k}\in\text{mBZ}}\sum_{\alpha,a,\eta,s}\!\left[\mathcal{B}_{\vex{R}}
	\frac{\left[V^{(\eta)}_{\alpha a}(\vex{k}+\!\vex{G})\right]^*}{\gamma}e^{-i\vex{k}\cdot\vex{R}}\bar{c}_{\vex{k}+\vex{G},a,\eta,s}f_{\vex{R},\alpha,\eta,s}\!+\!\mathcal{B}_{\vex{R}}^*\frac{V^{(\eta)}_{\alpha a}(\vex{k}+\!\vex{G})}{\gamma}e^{i\vex{k}\cdot\vex{R}}\bar{f}_{\vex{R},\alpha,\eta,s}c_{\vex{k}+\vex{G},a,\eta,s}\right]\!\!+\!\frac{|\mathcal{B}_{\vex{R}}|^2}{J_K}\!\right\}\!,\\
	\nonumber&-J_H\int\limits_{\tau}\sum_{\langle \vex{R},\vex{R}'\rangle}\sum_{\alpha,\alpha',\eta,\eta',s,s'}\bar{f}_{\vex{R},\alpha,\eta,s}f_{\vex{R}',\alpha,\eta,s}\bar{f}_{\vex{R}',\alpha',\eta',s'}f_{\vex{R},\alpha',\eta',s'}\\
	\rightarrow&\int\limits_{\tau}\sum_{\langle \vex{R},\vex{R}'\rangle}\left\{\sum_{\alpha,\eta,s}\left[\chi_{\vex{R},\vex{R}'}\bar{f}_{\vex{R}',\alpha,\eta,s}f_{\vex{R},\alpha,\eta,s}+\chi^*_{\vex{R},\vex{R}'}\bar{f}_{\vex{R},\alpha,\eta,s}f_{\vex{R}',\alpha,\eta,s}\right]+\frac{1}{J_H}\left|\chi_{\vex{R},\vex{R}'}\right|^2
	\right\},
\end{align}

Now, the partition function becomes
\begin{align}
	Z=&\int\mathcal{D}\left[c,\bar{c},f,\bar{f},\mathcal{B},\mathcal{B}^*,\chi,\chi^*;\mu_f,\mu_c\right]e^{-\mathcal{S}'},\\
	\nonumber\mathcal{S}'=&\int\limits_{\tau}\sum_{\eta,s}\sum_{a,a'}\sum_{\vex{q}}\bar{c}_{\vex{q},a,\eta,s}\left[\partial_{\tau}+\hat{h}^{(\eta)}_{a,a'}(\vex{q})-\mu_c\right]c_{\vex{q},a',\eta,s}+\int\limits_{\tau}\sum_{\vex{R}}\sum_{\alpha,\eta,s}\bar{f}_{\vex{R},\alpha,\eta,s}\left[\partial_{\tau}-\mu_f(\vex{R})\right]f_{\vex{R},\alpha,\eta,s}\\
	\nonumber&+\int\limits_{\tau}\sum_{\vex{R}}\frac{1}{\sqrt{\mathcal{N}_k}}\sum_{\vex{G}}\sum_{\vex{k}\in\text{mBZ}}\sum_{\alpha,a,\eta,s}\left[\mathcal{B}_{\vex{R}}
	\frac{\left[V^{(\eta)}_{\alpha a}(\vex{k}+\!\vex{G})\right]^*}{\gamma}e^{-i\vex{k}\cdot\vex{R}}\bar{c}_{\vex{k}+\vex{G},a,\eta,s}f_{\vex{R},\alpha,\eta,s}+\mathcal{B}_{\vex{R}}^*\frac{V^{(\eta)}_{\alpha a}(\vex{k}+\!\vex{G})}{\gamma}e^{i\vex{k}\cdot\vex{R}}\bar{f}_{\vex{R},\alpha,\eta,s}c_{\vex{k}+\vex{G},a,\eta,s}\right]\\
	\nonumber&+\int\limits_{\tau}\sum_{\langle \vex{R},\vex{R}'\rangle}\sum_{\alpha,\eta,s}\left[\chi_{\vex{R},\vex{R}'}\bar{f}_{\vex{R}',\alpha,\eta,s}f_{\vex{R},\alpha,\eta,s}+\chi^*_{\vex{R},\vex{R}'}\bar{f}_{\vex{R},\alpha,\eta,s}f_{\vex{R}',\alpha,\eta,s}\right]
	\\
	&+\int\limits_{\tau}\sum_{\vex{R}}\left(\frac{\left|\mathcal{B}_{\vex{R}}\right|^2}{J_K}+\mu_f(\vex{R})N_f\right)+\int\limits_{\tau}\sum_{\langle\vex{R},\vex{R}'\rangle}\frac{\left|\chi_{\vex{R},\vex{R}'}\right|^2}{J_H}+\beta\mu_c\mathcal{N}_kN_c.
\end{align}
The saddle point solutions correspond to
\begin{subequations}
	\begin{align}
		\frac{\delta \mathcal{S}'}{\delta \mathcal{B}_{\vex{R}}^*}=0\rightarrow&\frac{1}{\sqrt{\mathcal{N}_k}}\sum_{\vex{G}}\sum_{\vex{k}\in\text{mBZ}}\sum_{\alpha,\eta,s}\frac{V^{(\eta)}_{\alpha a}(\vex{k}+\vex{G})}{\gamma}e^{i\vex{k}\cdot\vex{R}}\left\langle\bar{f}_{\vex{R},\alpha,\eta,s}c_{\vex{k}+\vex{G},a,\eta,s}\right\rangle+\frac{\mathcal{B}_{\vex{R}}}{J_K}=0,\\
		\frac{\delta \mathcal{S}'}{\delta \chi_{\vex{R},\vex{R}'}^*}=0\rightarrow&\sum_{\alpha,\eta,s}\left\langle\bar{f}_{\vex{R},\alpha,\eta,s}f_{\vex{R}',\alpha,\eta,s}\right\rangle+\frac{\chi_{\vex{R},\vex{R}'}}{J_H}=0,\\
		\frac{\delta \mathcal{S}'}{\delta \mu_f(\vex{R})}=0\rightarrow&\sum_{\alpha,\eta,s}\left\langle\bar{f}_{\vex{R},\alpha,\eta,s}f_{\vex{R},\alpha,\eta,s}\right\rangle-N_f=0\\
		\frac{\delta \mathcal{S}'}{\delta \mu_c}=0\rightarrow&\sum_{\vex{G}}\sum_{\vex{k}\in \text{mBZ}}\sum_{a,\eta,s}\left\langle\bar{c}_{\vex{k}+\vex{G},a,\eta,s}c_{\vex{k}+\vex{G},a,\eta,s}\right\rangle-\mathcal{N}_kN_c=0
	\end{align}
\end{subequations}

For a fixed realization of $\mathcal{B}$, $\chi$, $\mu_f$, and $\mu_c$ the problem is quadratic in fermionic fields. Thus, we can formally integrate out fermions and construct an effective action. There are many possible choices of the static solutions corresponding to different phases. In particular, we are interested in exploring the interplay between Kondo hybridization and magnetic correlation. We consider a static mean-field ansatz as follows:
\begin{align}
	\mathcal{B}_{\vex{R}}=Be^{i\phi},\,\,\chi_{\vex{R},\vex{R}'}=\chi,\,\,\,\mu_f(\vex{R})=\mu_f
\end{align}
where $B<0$, $\chi>0$, $\mu_f$ is real-valued, and $\phi$ is the phase of hybridization. The phase $\phi$ can be gauged away by a unitary transformation, so we consider only the $\phi=0$ case without loss of generality. Now, $\mu_f$ is the chemical potential for $f$ fermions. 
Before we proceed, we discuss the $f$ fermion sector with the mean field ansatz. In the absence of hybridization, the ground state with $\chi_{\vex{R},\vex{R}'}=\chi$ realizes a spin liquid with a spinon Fermi surface. This might not be the true competing order against the Kondo semimetal, but this choice is symmetry-preserving and translational invariant on the triangular lattice.

With the mean-field and saddle-point approximations, the imaginary-time action is given by
\begin{align}
	\mathcal{S}_{\text{MF}}=&\int\limits_{\tau}\sum_{\eta,s}\sum_{a,a'}\sum_{\vex{q}}\bar{c}_{\vex{k},a,\eta,s}\left[\partial_{\tau}+\hat{h}^{(\eta)}_{a,a'}(\vex{q})-\mu_c\right]c_{\vex{q},a',\eta,s}+\int\limits_{\tau}\sum_{\vex{R}}\sum_{\alpha,\eta,s}\bar{f}_{\vex{R},\alpha,\eta,s}\left[\partial_{\tau}-\mu_f\right]f_{\vex{R},\alpha,\eta,s}\\
	\nonumber&+\frac{B}{\gamma}\int\limits_{\tau}\sum_{\vex{R}}\frac{1}{\sqrt{\mathcal{N}_k}}\sum_{\vex{G}}\sum_{\vex{k}\in\text{mBZ}}\sum_{\alpha,a,\eta,s}\left\{
	\left[V^{(\eta)}_{\alpha a}(\vex{k}+\!\vex{G})\right]^*e^{-i\vex{k}\cdot\vex{R}}\bar{c}_{\vex{k}+\vex{G},a,\eta,s}f_{\vex{R},\alpha,\eta,s}+V^{(\eta)}_{\alpha a}(\vex{k}+\!\vex{G})e^{i\vex{k}\cdot\vex{R}}\bar{f}_{\vex{R},\alpha,\eta,s}c_{\vex{k}+\vex{G},a,\eta,s}\right\}\\
	\nonumber&+\chi\int\limits_{\tau}\sum_{\langle \vex{R},\vex{R}'\rangle}\sum_{\alpha,\eta,s}\left[\bar{f}_{\vex{R}',\alpha,\eta,s}f_{\vex{R},\alpha,\eta,s}+\bar{f}_{\vex{R},\alpha,\eta,s}f_{\vex{R}',\alpha,\eta,s}\right]
	\\
	&+\beta\mathcal{N}_k\left(\frac{B^2}{J_K}+\frac{3\chi^2}{J_H}+\mu_fN_f+\mu_cN_c\right)\\
	=&\frac{1}{\beta}\sum_{\omega_n}\sum_{\vex{k}\in\text{mBZ}}\sum_{\eta,s}\hat{\bar{\Psi}}^{\eta,s}_{\omega_n,\vex{k}}\left[-i\omega_n\hat{1}_{(2+4N_G)\times (2+4N_G)}+\hat{\mathcal{H}}^{\eta,s}(\vex{k};B,\chi,\mu_f,\mu_c)\right]\hat{\Psi}^{\eta,s}_{\omega_n,\vex{k}}
	+\beta\mathcal{N}_k\left(\frac{B^2}{J_K}+\frac{3\chi^2}{J_H}+\mu_fN_f+\mu_cN_c\right),
\end{align}
where $\hat{\Psi}$ is a $(2+4N_G)$-component field made of the $f$ as well as $c$ fermions,
\begin{align}
	\hat{\mathcal{H}}^{\eta,s}(\vex{k};B,\chi,\mu_f,\mu_c)=&\left[\begin{array}{cc}
		(-\mu_f+\chi\epsilon_{\vex{k}})\hat{1}_{2\times 2} & \frac{B}{\gamma}\hat{\mathcal{V}}^{(\eta)}(\vex{k})\\
		\frac{B}{\gamma}\left[\hat{\mathcal{V}}^{(\eta)}(\vex{k})\right]^{\dagger} & \hat{\mathcal{H}}^{(\eta)}_c(\vex{k})
	\end{array}
	\right],\\
	\epsilon_{\vex{k}}=&2\cos\left(k_ya_M\right)+2\cos\left(\frac{\sqrt{3}}{2}k_xa_M+\frac{1}{2}k_ya_M\right)+2\cos\left(\frac{\sqrt{3}}{2}k_xa_M-\frac{1}{2}k_ya_M\right),
\end{align}
$\vex{r}_1=(0,1)a_M$, $\vex{r}_2=(\sqrt{3}/2,1/2)a_M$, and $\vex{r}_3=(\sqrt{3}/2,-1/2)a_M$,
and $a_M$ is the moir\'e lattice constant. $\epsilon_{\vex{k}}$ corresponds to the dispersion of a hopping Hamiltonian on a triangular lattice. The self-consistent equations become
\begin{subequations}\label{Eq:SCE_S}
	\begin{align}
		&B\!=\!-\frac{J_K}{\mathcal{N}_k}\!\sum_{\vex{G}}\sum_{\vex{k}\in\text{mBZ}}\sum_{\alpha,\eta,s}\!\frac{V^{(\eta)}_{\alpha a}(\vex{k}\!+\!\vex{G})}{\gamma}\!\left\langle f^{\dagger}_{\vex{k},\alpha,\eta,s}c_{\vex{k}+\vex{G},a,\eta,s}\right\rangle,\\
		&\chi=-\frac{J_H}{3\mathcal{N}_k}\sum_{n=1}^3\sum_{\vex{k}\in\text{mBZ}}\sum_{\alpha,\eta,s}e^{i\vex{k}\cdot\vex{r}_n}\left\langle\bar{f}_{\vex{k},\alpha,\eta,s}f_{\vex{k},\alpha,\eta,s}\right\rangle,\\
		&N_f=\frac{1}{\mathcal{N}_k}\sum_{\vex{k}\in\text{mBZ}}\sum_{\alpha,\eta,s}\left\langle\bar{f}_{\vex{k},\alpha,\eta,s}f_{\vex{k},\alpha,\eta,s}\right\rangle,\\
		&N_c=\frac{1}{\mathcal{N}_k}\sum_{\vex{G}}\sum_{\vex{k}\in\text{mBZ}}\sum_{a,\eta,s}\left\langle\bar{c}_{\vex{k}+\vex{G},a,\eta,s}c_{\vex{k}+\vex{G},a,\eta,s}\right\rangle
	\end{align}
\end{subequations}

Now, we formally integrate out fermionic fields at the level of partition function and derive the effective action as follows:
\begin{align}
	\mathcal{S}_{\text{eff}}[B,\chi,\mu_f,\mu_c]=&-\sum_{\omega_n}\sum_{\vex{k}\in\text{mBZ}}\sum_{\eta,s}\ln\det\left[-i\omega_n+\hat{\mathcal{H}}^{\eta,s}(\vex{k};B,\chi,\mu_f,\mu_c)\right]+\beta\mathcal{N}_k\left(\frac{\mathcal{B}^2}{J_K}+3\frac{\chi^2}{J_H}+\mu_f N_f+\mu_cN_c\right)\\
	=&-\sum_{\omega_n}\sum_{\vex{k}\in\text{mBZ}}\sum_{\eta,s}\sum_{b=1}^{2+4N_G}\ln\left[-i\omega_n+\mathcal{E}^{\eta,s}_b(\vex{k};B,\chi,\mu_f,\mu_c)\right]+\beta\mathcal{N}_k\left(\frac{\mathcal{B}^2}{J_K}+3\frac{\chi^2}{J_H}+\mu_f N_f+\mu_cN_c\right),
\end{align}
where $\mathcal{E}^{\eta,s}_b(\vex{k};B,\chi,\mu_f,\mu_c)$ is the eigenvalue of $\hat{\mathcal{H}}^{\eta,s}(\vex{k};B,\chi,\mu_f,\mu_c)$.
The mean-field free energy per site is given by
\begin{align}
	\frac{F_{\text{MF}}}{\mathcal{N}_k}=&\frac{\mathcal{S}_{\text{eff}}}{\beta\mathcal{N}_k}=-\frac{1}{\beta \mathcal{N}_k}\sum_{\omega_n}\sum_{\vex{k}\in\text{mBZ}}\sum_{\eta,s}\sum_{b=1}^{2+4N_G}\ln\left[-i\omega_n+\mathcal{E}^{\eta,s}_b(\vex{k};B,\chi,\mu_f,\mu_c)\right]+\left(\frac{\mathcal{B}^2}{J_K}+3\frac{\chi^2}{J_H}+\mu_f N_f+\mu_cN_c\right)\\
	=&-\frac{1}{\beta\mathcal{N}_k}\sum_{\vex{k}\in\text{mBZ}}\sum_{\eta,s}\sum_{b=1}^{2+4N_G}\ln\left[1+e^{\beta \mathcal{E}^{\eta,s}_b(\vex{k};B,\chi,\mu_f,\mu_c)}\right]+\left(\frac{\mathcal{B}^2}{J_K}+3\frac{\chi^2}{J_H}-\lambda N_f+\mu_cN_c\right),\\
	\lim\limits_{T\rightarrow 0}\frac{F_{\text{MF}}}{\mathcal{N}_k}=&\frac{E_{\text{MF}}}{\mathcal{N}_k}=\frac{1}{\mathcal{N}_k}\sum_{\vex{k}\in \text{mBZ}}\sum_{\eta,s}\!\sum_{b=1}^{2+4N_G}\mathcal{E}^{\eta,s}_b(\vex{k})\Theta(-\mathcal{E}^{\eta,s}_b(\vex{k}))+\frac{1}{\Omega_0}\left(\frac{\mathcal{B}^2}{J_K}+3\frac{\chi^2}{J_H}+\mu_f N_f+\mu_cN_c\right).
\end{align}
To derive the mean-field energy ($E_{\text{MF}}$), we have used $\lim\limits_{\beta\rightarrow \infty}\left[-\beta^{-1}\ln(1+e^{\beta E})\right]\rightarrow E\Theta(-E)$ with $\Theta(x)$ being the Heaviside function.

\section{Evolution of quasiparticle dispersion with different $|B|$ for $\nu=0$}

\begin{figure}[t!]
	\includegraphics[width=0.8\textwidth]{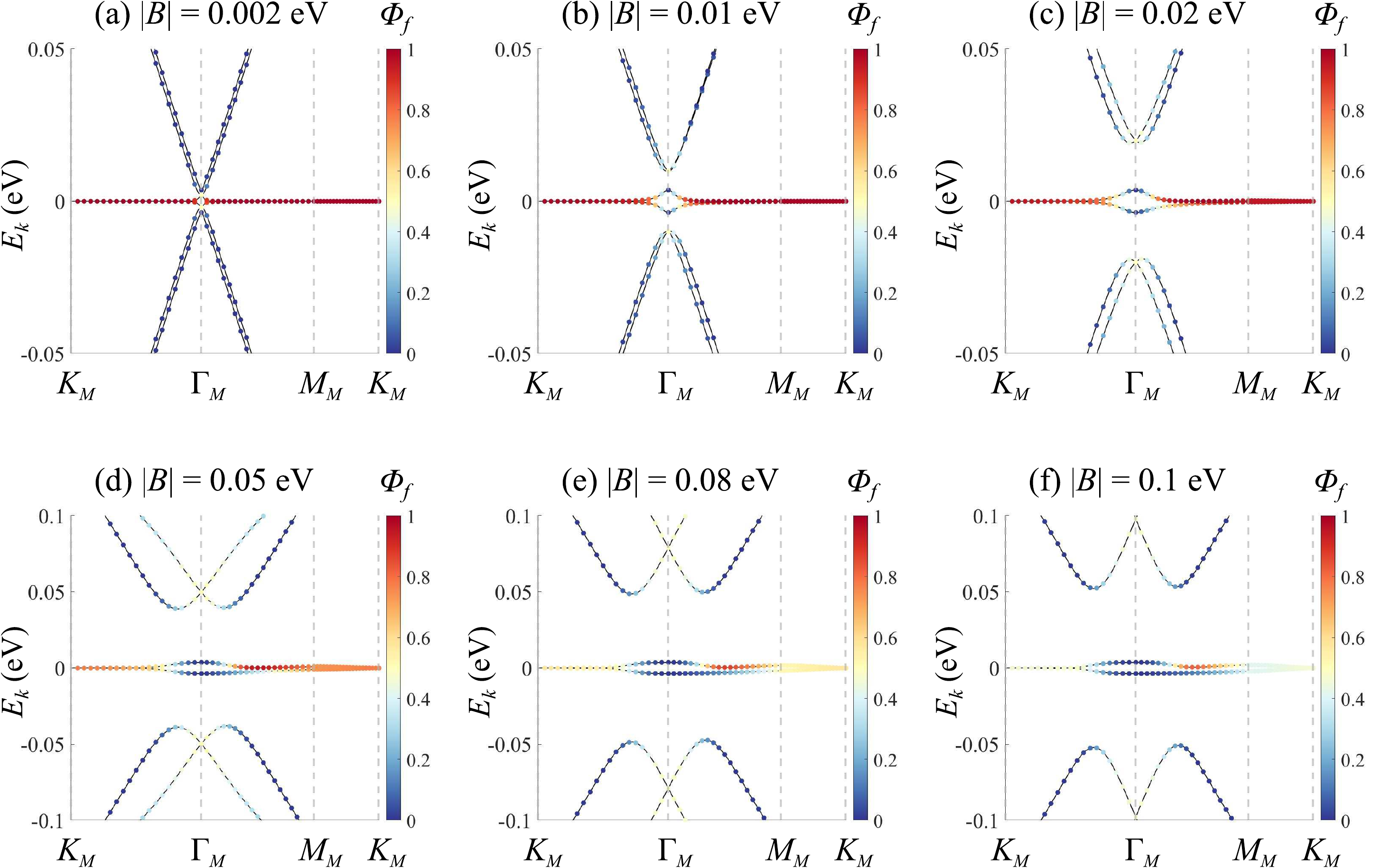}
	\caption{Kondo semimtal dispersion ($\nu=0$) with different values of Kondo hybridization amplitude $|B|$. The color represents the composition of $f$ fermion $\Phi_f$ in the corresponding wavefunction. $\Phi_f=1$ (red) means the wavefucntions are made of $f$ fermions only; $\Phi_f=0$ (blue) indicates the wavefucntions are made of $c$ fermions only. $\eta=+1$ for all the plots.
	}		
	\label{Fig:KSM}
\end{figure}

It is interesting to investigate how the hybridization amplitude affects the quasiparticle dispersion in the Kondo semimetals, corresponding to $\nu=\nu_c=\nu_f=0$. As we discuss in the main text, a finite Kondo hybridization (i.e., a nonzero $|B|$) forms for $J_K>0.01$ eV. In Fig.~\ref{Fig:KSM}, we plot the quasiparticle dispersion with a few representative values of $|B|$, demonstrating the evolution of quasiparticle dispersion. In Fig.~\ref{Fig:KSM}(a), the localized $f$ fermions and the delocalized $c$ fermions are nearly decoupled except for a few states near $\Gamma_M$. In such as situation, the mini bands are not isolated from the nearby bands. The dispersion becomes qualitatively similar to the single-particle MATBG bands for $|B|\ge M\approx 0.0037$ eV. In Fig.~\ref{Fig:KSM}(b)-(f), the mini bands form fragile topological Dirac Kondo semimetals, and the bandwidth is around $2M$, the same as the single-particle case. The single-particle case is close to $|B|=0.02$ eV [Fig.~\ref{Fig:KSM}(c)]. The interaction effect is encoded in the band gaps between the mini and remote bands. Moreover, $\Phi_f$ becomes smaller in the mini bands for a larger $J_K$, indicating strong band reconstruction.

\section{Quasiparticle dispersion for nonzero integer fillings}

\begin{figure}[t!]
	\includegraphics[width=0.8\textwidth]{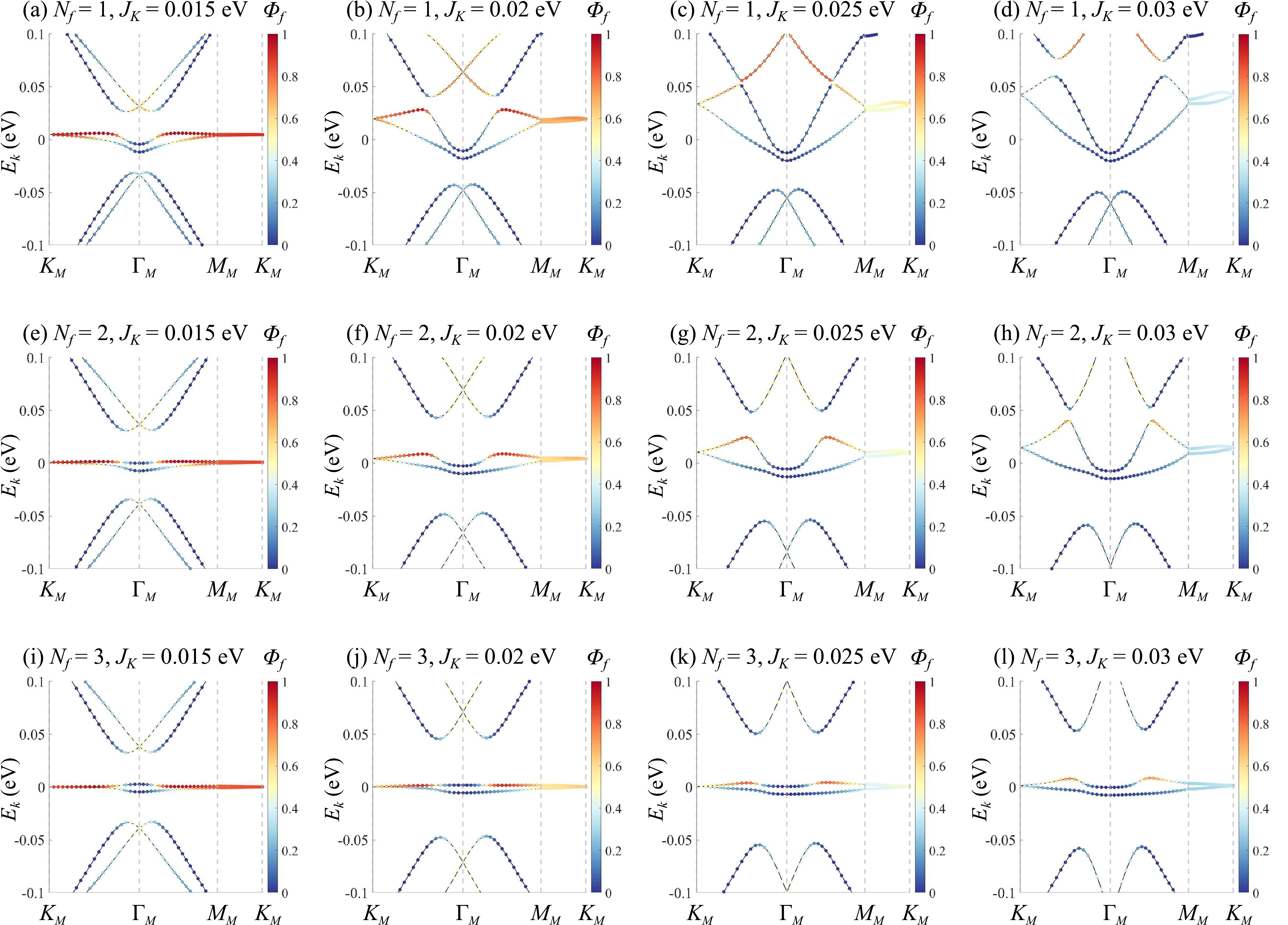}
	\caption{Quasiparticle dispersion for $N_f=1,2,3$ (corresponding to $\nu=-3,-2,-1$.). (a)-(d): $N_f=1$; (e)-(f): $N_f=2$; (i)-(l): $N_f=3$. We plot the quasiparticle bands by solving the self-consistent equations with $J_K=0.015,0.02,0.025,0.03$ eV. The color represents the composition of $f$ fermion $\Phi_f$ in the corresponding wavefunction. $\eta=+1$ for all the plots.
	}		
	\label{Fig:KSM_Nf}
\end{figure}

In the main text, we reveal the broadening of the quasiparticle low-energy bands for $N_f\neq 4$. Here, we briefly discuss the quasiparticle dispersion with different $N_f$ and $J_K$. In Fig.~\ref{Fig:KSM_Nf}, we plot quasiparticle dispersion with different $N_f$ and $J_K$ (equivalent to $B$ after solving the self-consistent equations). First, we notice that the low-energy bands become wider for a larger $J_K$ for the range of $J_K$ presented in Fig.~\ref{Fig:KSM_Nf}. For a sufficiently small $J_K$ corresponding to $|B|<M$, the $f$ orbitals and the delocalized bands are almost decoupled, similar to the dispersion of Fig.~\ref{Fig:KSM_Nf}(a). Furthermore, the value of $N_f$ also significantly affects the broadening of the low-energy band. In particular, the low-energy bandwidth gets wider for a larger $|N_f-4|$. On the other hand, the value of the Kondo hybridization amplitude $|B|$ shows the opposite trend, i.e., $N_f=4$ case has the largest $|B|$, while the $N_f=1$ case gives the smallest $|B|$. The above observations imply that $|B|$ alone is insufficient to determine the broadening of the low-energy bands. To understand the low-energy bands, one has to consider $\mu_f$ and $\mu_c$, which are also obtained from solving the self-consistent equations. Intuitively, a finite $|\mu_f-\mu_c|$ corresponds to coupling local moments with a finite Fermi surface of $c$ fermions, and the low-energy bands can become quite dispersive. This situation is very different from the band reconstruction of the single-particle Song-Bernevig model \cite{Song2022_S}, where the Fermi surface is zero. We also note that the interaction terms we ignore in our calculations may modify the value of $|\mu_f-\mu_c|$, causing quantitative changes of band structures. Physically, the nonzero $|\mu_f-\mu_c|$ is a consequence of interaction.

We also point out that phase transitions exist in the quasiparticle bands of $N_f=1$ ($\nu=-3$) and $N_f=2$ ($\nu=-2$). For $N_f=1$, accidental band touching between the upper mini band and the higher energy band takes place near $J_K\approx 0.024$ eV. The similar band touching also happens around $J_K\approx0.036$ eV for $N_f=2$.

\section{Numerical procedures for solving self-consistent equations}

We diagonalize the $(2+N_G)\times (2+N_G)$ matrix $\hat{H}^{\eta,s}(\vex{k};B,\chi,\mu_f,\mu_c)$ with initial values of $B,\chi,\mu_f,\mu_c$ and evaluate the left hand sides of Eq.~(\ref{Eq:SCE_S}). Then, we update $\mathcal{B}$, $\chi$, $\mu_f$, and $\mu_c$ for the next iteration until $e<0.01$ where
\begin{align}
	e=\sqrt{\frac{|\delta B|^2}{|B|^2}+\frac{|\delta \chi|^2}{|\chi|^2}+\frac{\delta N_f^2}{N_f^2}+\delta N_c^2}.
\end{align}
We iterate the self-consistent equations with $N_G=7$. Then, we repeat the numerical calculations with $N_G=37$ using the solutions from the $N_G=7$ calculations. For most cases, esults with $ N_G=7 $ and $ N_G=37 $ are identical within our numerical accuracy. Finally, we check $E_{\text{MF}}$ and pick the configuration minimize $E_{\text{MF}}$.

\section{Finite-size analysis}

\begin{figure}[h!]
	\includegraphics[width=0.4\textwidth]{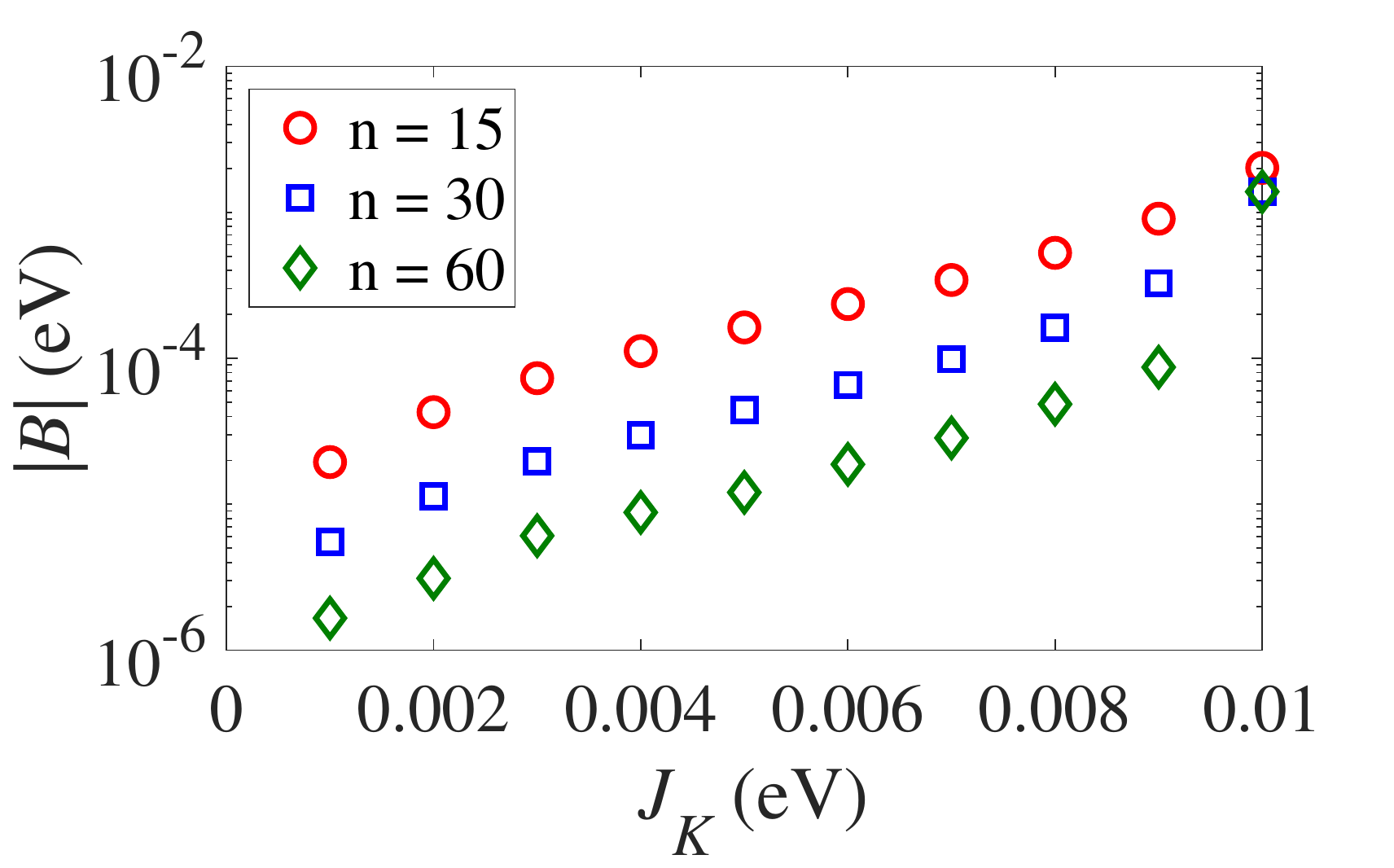}
	\caption{The Kondo hybridization amplitude for $N_f=4$ with different sizes of momentum mesh. We consider a $n\times n$ momentum mesh for each mBZ.
	}		
	\label{Fig:Finite_n}
\end{figure}

In all the calculations, we consider a $n\times n$ momentum mesh for each mBZ with $n=30$. We find that most of the results converge for $n\ge 30$. The only exception is the Kondo lattice calculations with $J_K<0.01$ eV. In Fig.~\ref{Fig:Finite_n}, we plot the Kondo hybridization amplitude $|B|$ for $N_f=4$ with different $n$. The results show that $|B|$ decreases as $n$ increases, suggesting that $|B|$ vanishes in the thermodynamic limit. We also find similar results for $N_f=1,2,3$. As such, we conclude that there is a finite threshold (with $J_K\approx 0.01$ eV) for the Kondo hybridization formation in our Kondo lattice models.

\begin{thebibliography}{67}
\bibitem[{NO()}]{NO}
	\bibinfo{note}{The normal ordering is given by: $:\hat{A}:\equiv A-\langle
		G|A|G\rangle$, where $\langle
		G|c^{\dagger}_{\vex{q},a,\eta,s}c_{\vex{q},a',\eta',s'}|G\rangle=\frac{1}{2}\delta_{\vex{q},\vex{q}'}\delta_{a,a'}\delta_{\eta,\eta'}\delta_{s,s'}$,
		$\langle
		G|f^{\dagger}_{\vex{k},\alpha,\eta,s}f_{\vex{k},\alpha',\eta',s'}|G\rangle=\frac{1}{2}\delta_{\vex{k},\vex{k}'}\delta_{\alpha,\alpha'}\delta_{\eta,\eta'}\delta_{s,s'}$,
		and $\langle
		G|f^{\dagger}_{\vex{k},\alpha,\eta,s}c_{\vex{q},a',\eta',s'}|G\rangle=0$.}
\bibitem[{KSM()}]{KSM}
	\bibinfo{note}{For $|B|<3.7$ meV, the minibands are not isolated. See \cite{SM}
		for a discussion.}
\end{thebibliography}

\end{document}